# Deep Pareto Reinforcement Learning for Multi-Objective Recommender Systems

Pan Li[1], Alexander Tuzhilin[2]

[1]Scheller College of Business, Georgia Tech, pli95@gatech.edu

[2]Stern School of Business, New York University, at2@stern.nyu.edu

**Abstract**

Optimizing multiple objectives simultaneously is an important task for recommendation platforms to improve their performance. However, this task is particularly challenging since the relationships between different objectives are heterogeneous across different consumers and dynamically fluctuate according to different contexts, resulting in a Pareto-frontier in the result of recommendations, where the improvement of any objective comes at the cost of others. Existing multi-objective recommender systems do not systematically consider such dynamic relationships; instead, they balance between these objectives in a static and uniform manner, resulting in only suboptimal recommendation performance. In this paper, we propose a *Deep Pareto Reinforcement Learning* (DeepPRL) method, where we (1) comprehensively model the complex relationships between multiple recommendation objectives; (2) effectively capture personalized and contextual consumer preferences for each objective; (3) optimize both the short-term and the long-term recommendation performance. As a result, our method achieves significant Pareto-dominance over the state-of-the-art baselines across four offline experiments. Furthermore, we conducted a controlled experiment on Alibaba's video streaming platform, where our method simultaneously improved three conflicting business objectives significantly over the latest production system, demonstrating its tangible economic impact in practice.

## 1 Introduction

Multi-objective recommender systems (Adomavicius et al. 2010) have experienced widespread adoption in major companies, such as YouTube (Ma et al. 2018) and Alibaba (Ma et al. 2018b), to improve the business performance across multiple fronts, e.g., maximizing revenues from ads *and* improving click-through rates from consumers (Kamakura et al. 1996; Roijers et al. 2013). One critical challenge is associated with the Pareto frontier issue, where improving the value of one objective may come at the expense of another (Ribeiro et al. 2014). For example, there is a well-known conflict between the objectives of Video View (VV, the average number of videos watched in each session) and Dwell Time (DT, the average time spent in each session) in video streaming platforms (Cheng et al. 2013). On one hand, focusing on VV may produce short-length, clickbait videos that reduce DT. On the other hand, if the focus is on DT, the platform tends to recommend long videos, which might not fit well with consumer satisfaction and lead to a decrease in VV. Since both objectives significantly impact consumer experiences (Sahoo et al. 2012), optimizing the Pareto frontier of recommendations becomes a crucial task.

Unfortunately, while multi-objective recommendation models have been proposed in the literature (Ribeiro et al. 2014), they do not systematically model the complex and dynamic relationships between objectives to optimize the Pareto frontier directly. Instead, static objective weights have been predominately used (Zheng & Wang 2022), resulting in only sub-optimal performance. This is the case, since the behavioral marketing theory (Häubl & Trifts 2000; Häubl et al. 2010) reveals that consumers' decision-making is contingent on the intrinsic consumer nature and the decision context, rather than being invariant. Therefore, their preferences for different objectives fluctuate accordingly. For example, some consumers prefer to stay within their own comfort zones to receive familiar products, while others are willing to enjoy relevant content and explore novel content at the same time. In the former case, relevance and novelty work exclusively in a conflicting manner, while they are complementary in the latter case. In another example, a certain consumer might be more willing to listen to rock-and-roll music on a weekday morning to be energized for the day, versus relaxing music on weekend afternoons.

To address these gaps, we propose a *Deep Pareto Reinforcement Learning (DeepPRL)* model in this paper, which continuously calibrates the objective relationships based on context and dynamic factors, rather than modeling them statically. It consists of the following two components that we propose in the paper: (1) "Mixture of HyperNetwork", in which we build multiple hypernetworks (Navon et al. 2021) to capture the latent information related to each objective, so that we can select suitable products to improve the Pareto frontier with respect to that objective. The outputs are aggregated through a mixture attention network that captures the intrinsic relationships between different objectives. (2) "Deep Contextual Reinforcement Learning", in which we dynamically adjust the weights of multiple objectives based on context and dynamic factors, leading to optimized performance along multiple fronts of the Pareto frontier. It also enables us to model the impact of current consumer actions on future consumer preferences (Afsar et al. 2022). Integration of these two modules into the unified DeepPRL model provides the following advantages. First, it effectively balances different objectives based on contextualized and evolving consumer preferences. Second, it improves both the short-term and the long-term performance, which further enhances its practicality, since many existing methods suffer degradation of performance in the long run (Devooght & Bersini 2017). Third, it provides theoretical properties that guarantee its superior performance, as explained in Section 4.

These advantages of DeepPRL are further validated through extensive offline experiments in four different applications of Alibaba-Youku, Yelp, Spotify, and Kuaishou, where our method significantly improves performance *across all considered objectives* over state-of-the-art baselines and dominates their Pareto frontiers, regardless of specific datasets or heterogeneous relationships between the objectives. We have also conducted a simulation study to illustrate the significant long-term performance improvements of DeepPRL. In addition, to demonstrate its practical impact, we conducted a large-scale online experiment at the video streaming platform of Alibaba, where DeepPRL *simultaneously* improved the performance of the three conflicting objectives of Click-Through Rate, Video View, and Dwell Time by 2%, 5%, and 7% respectively over the latest production system, thus producing tangible economic benefits for the platform.

In this paper, we make the following research contributions. First, we enhance the level of personalization and contextualization in multi-objective recommendations by selecting the appropriate set of objective weights for *each* consumer under *each* context, resulting in superior performance. Second, we propose a DeepPRL method that automatically and effectively identifies consumer preference towards different objectives and updates the objective weights in real-time, leading to performance improvements both in the short term and in the long run. Third, we present the theoretical properties of our method and empirically demonstrate its advantages vis-à-vis the state-of-the-art baselines through offline experiments, where it achieves significant improvements across multiple objectives and experimental settings, and dominates their Pareto-frontiers. Finally, we illustrate its practical impact through an online experiment, where it significantly outperforms the latest production system across three conflicting objectives.

## 2 Related Work

### 2.1 Multi-Objective Recommender System

As illustrated in the literature (Sahoo et al. 2012; Ribeiro et al. 2014), it is beneficial to optimize multiple, as opposed to only one, objectives in recommendations (Adomavicius & Tuzhilin 2005). This finding is justified in (Bauman & Tuzhilin 2022) and (Panniello et al. 2016), as consumers actively pursue multiple objectives to improve their experiences across different dimensions (Sahoo et al. 2012; Zimmermann et al. 2018). The economic benefits of multi-objective recommendations have also been discussed in terms of quality, taste-matching, diversity, and transparency in the literature (Xu et al. 2014; Shi & Raghu 2020; Yin et al. 2022).

To this end, two groups of multi-objective recommender systems have been proposed (Zheng & Wang 2022): The first group utilizes statistical learning methods, such as cross-stitch network (Misra et al. 2016), relational network (Zhao et al. 2019), and structural models (Panniello et al. 2016) to jointly predict the objective values (Adomavicius et al. 2010; Rodriguez et al. 2012). The second group develops novel network architectures, including Shared-Bottom (Ruder 2017), Multi-Gate Mixture-of-Expert (MMOE) (Ma et al. 2018), and Mixture of Sequential Experts (MoSE) (Qin et al. 2020), to predict the values of one specific objective and then aggregate these predicted objective values through a set of fixed weights to produce multi-objective recommendations. As a result, they manage to facilitate accurate joint predictions of multiple objectives and achieve strong performance in practice.

### 2.2 Personalized and Contextualized Consumer Preferences in Recommendations

Personalization technique enables marketers to reach their potential customers effectively (Ho & Lim 2018; Zhang et al. 2019) and produce customized experiences (Thirumalai & Sinha 2013), resulting in important competitive advantages (Johar et al. 2014). Therefore, personalization methods are widely adopted to identify suitable content (Adomavicius & Tuzhilin 2005), improve customer loyalty (Tongxiao et al. 2011), post-consumption experience (Adomavicius et al. 2021), and advertising effectiveness (Bleier & Eisenbeiss 2015). These methods typically design neural networks to identify personalized interests, such as Deep Reinforcement Network (DRN) (Zheng et al. 2018) and Deep Interest Network (DIN) (Zhou et al. 2018).

Another important technique is contextualization (Gorgoglione et al. 2019; Bauman & Tuzhilin 2022), which captures the evolution of consumer preference affected by contextual factors, such as prior activities, time, and location (Palmisano et al. 2008). As shown in the literature (Panniello et al. 2016), contextualization significantly improves the accuracy and diversity of recommendations, which in turn improves business performance, such as sales and retention rates. Some representative contextualization methods include Hierarchical Linear Models (Umyarov & Tuzhilin 2009) and BERT4Rec (Sun et al. 2019).

### 2.3 Pareto-Oriented Recommender System

Our work is related to the work that approximates Pareto optimality in recommendations. For example, researchers in (Chen et al. 2021) developed a Pareto optimization method by incorporating a gradient-based model with the actor-critic learning process. In (Jin et al. 2024), researchers introduced a Pareto optimization solver to facilitate a better objective tradeoff, and achieved favorable outcomes on WeChat Channels. Furthermore, (Lin et al. 2019) presented a general framework for producing Pareto efficient recommendations, assuming that there are formal differentiable formulations for all objectives, while (Xie et al. 2021) removed this requirement and developed personalized weights for different users to achieve approximated Pareto efficient performance. Finally, researchers also proposed multiple Pareto-oriented models for tackling the fairness constraints in recommender systems (Hao et al. 2021, Ge et al. 2022).

**2.4 Research Gaps In the Literature**

There exist several problems with all three streams of prior work. First, regarding existing multi-objective recommender systems, they do not consider heterogeneous and dynamically evolving relationships between objectives, which is a crucial factor in providing effective multi-objective recommendations, as illustrated in the next section. In addition, they do not systematically model the personalized and contextualized consumer preference towards different objectives, which plays an important role in multi-objective recommendations. Second, for existing personalized or contextualized methods, they primarily focus on optimizing a single objective (such as Click-Through Rate), while the extension to a multi-objective scenario is not trivial. Third, prior Pareto-oriented recommender systems do not provide any theoretical properties of the performance levels, and it is unclear whether they will be close to Pareto optimality or not. To address these gaps, we propose a novel DeepPRL method to model evolving relationships of different objectives, leading to superior performance both theoretically and empirically.

## 3 Preliminaries and Model-Free Evidence

### 3.1 Problem Formulation

In this section, we will present the model-free evidence to demonstrate the importance of context and dynamic factors in multi-objective recommendations. We first formulate the learning task.

We denote recommendation objectives as $\{y_1, y_2 \cdots, y_n\}$, where $n$ is the number of objectives, and each $y_i$ represents a particular objective (e.g., novelty, relevance). In many cases, however, the values of these objectives are not available prior to recommendation. For example, VV and DT can only be obtained from consumer responses after the video is recommended. Therefore, *a priori*, we need to predict their values $\{\widetilde{y_1}, \widetilde{y_2} \cdots, \widetilde{y_n}\}$ based on consumer and product features. Recommendations will then be produced by maximizing the utility value $\sum_k \alpha_k \widetilde{y_k}$, where each $\alpha_k$ represents the weight of objective $y_k$. The weighted sum utility function design is motivated by industrial practices, where major companies such as Google (Ma et al. 2018) and Amazon (Carmel et al. 2020) have all adopted such a utility in their systems. Since these companies have conducted numerous online A/B tests to explore the design that aligns best with their business interests, we simply follow their standard protocols in this paper. Therefore, we produce multi-objective recommendations by solving the following two-stage optimization problem:

$$(1)\ min \sum_k |\widetilde{y_k} - y_k| \quad \text{(Stage 1: Minimizing the Prediction Error of Each Objective)}$$

$$(2)\ max \sum_{k} \alpha_k \widetilde{y_k} \qquad \text{(Stage 2: Identifying the Optimal Weight of Each Objective)}$$

This optimization is particularly challenging since it is an NP-hard problem (Xiao et al. 2017). In addition, existing solutions focus primarily on the first stage to predict $\widetilde{y_k}$, while the second stage has been oversimplified in the sense that $\alpha_k$ is selected as a fixed value across different consumers and contexts (Ruder 2017; Ma et al. 2018; Qin et al. 2020), resulting in suboptimal performance. In the rest of this section, we will explain that both optimization tasks are heavily influenced by context and dynamic factors, which motivates the design of our DeepPRL model.

### 3.2 Motivating Example – Personalization & Contextualization

We will begin with a motivating example based on a dataset collected from the video streaming platform of Alibaba, with over 26 million video-watching records from July 2019. Videos are recommended using the latest production system to optimize a wide range of objectives, including Video View (VV) and Dwell Time (DT). As we explained in Section 1, these two objectives are conflicting with each other in general. However, the level of negative correlations can fluctuate dramatically with respect to different consumers and contexts, as shown below.

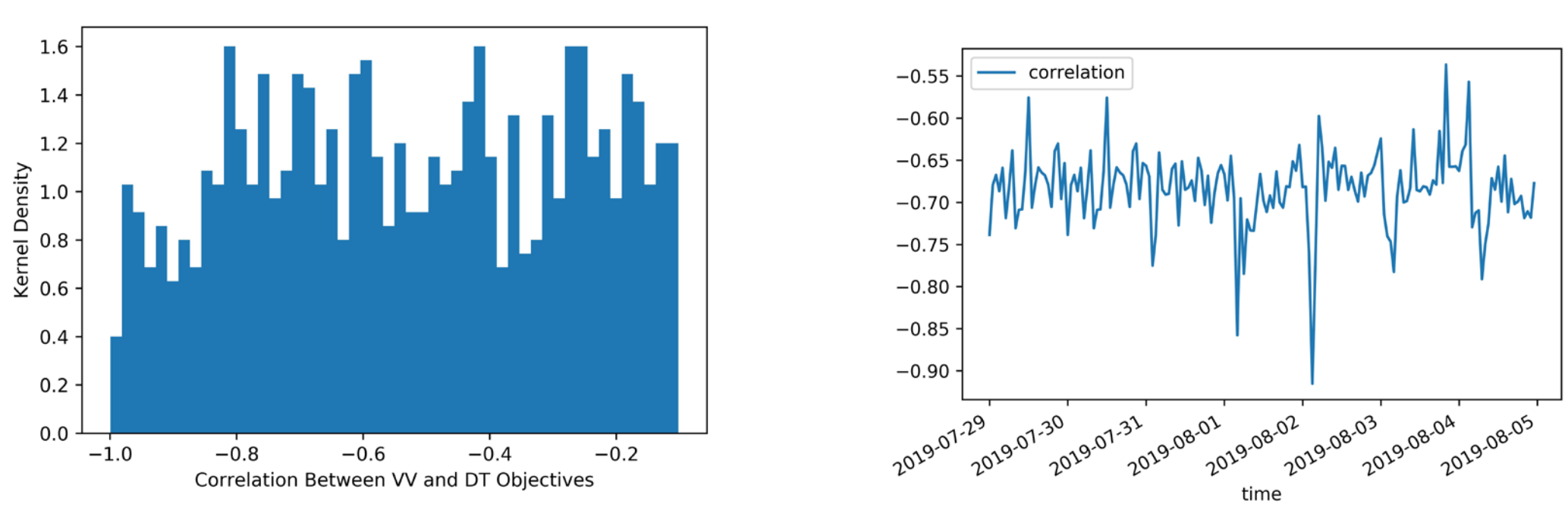

*(a) Different Consumers* *(b) Different Contexts of Time*

*Figure 1: Visualization of Correlations Between VV and DT across Different Scenarios*

**Evidence of Personalization.** We compute the Pearson Correlation between VV and DT

individually for each consumer and present the histogram in Figure 1(a), where the x-axis shows the correlation coefficients, and the y-axis represents the kernel density. We observe from the figure that the relationship is highly heterogeneous: although the average correlation is -0.663, 29% of consumers have a correlation below -0.7, and 47% of consumers have a correlation above -0.5. Since there are significant variations across different consumers, it is crucial to capture the *personalized* information, such as demographics and characteristics of consumers, to model their heterogeneous preferences towards different objectives in recommendations.

**Evidence of Contextualization.** Context is defined as "any additional information besides users and items that is relevant to recommendations" (Adomavicius et al. 2021b). We consider the context of the date and hour of video watching in our example, and compute the Pearson Correlation between VV and DT across different times in our dataset. We similarly present the average correlation levels in Figure 1(b), where we observe that the relationship between them fluctuates significantly across different time periods, thus further demonstrating the importance of properly modeling contextual information, such as time, in multi-objective recommendations.

### 3.3 Optimization of the Pareto Frontier

Building on the empirical evidence from Section 3.2, we will now present a toy example to show that incorporating personalized and contextualized information would significantly improve the Pareto frontier of multi-objective recommendations. We first introduce the preliminaries.

In multi-objective optimization, the Pareto optimality refers to the set of solutions where there is no way to make one objective better off without making others worse off (Emmerich & Deutz, 2018). To achieve such a performance level, multi-task learning is commonly adopted (Ruder 2017) to simultaneously solve several learning tasks (one for each objective) by sharing information between them. The final solutions are obtained through Linear Scalarization (LS) (Zhang & Yang 2021), in which each objective's weight is chosen a priori in a fixed manner. To capture the evolving relationships between different tasks, some recent attempts have been made, including gradient magnitude (Chen et al., 2018), attention (Liu et al. 2019), and task uncertainty (Kendall et al. 2018). However, these methods do not systematically model the trade-off between different objectives and the dynamic factors; as a result, they achieve only suboptimal performance that is significantly worse than the theoretical Pareto optimal.

In the following toy example, we provide multi-objective recommendations for two objectives $o_1$ and $o_2$, which are conflicting with each other, but are both beneficial to the consumer experience. Following existing practices (Sun et al. 2019b), we simulate the perceived value of objectives 1 & 2 for user $u$ and item $i$ using the discrete choice model as:

$$o_{u,i,1} = L\left[r_i + \eta_{u,i}\right] \qquad o_{u,i,2} = 1 - c_{u,i} \times o_{u,i,1}$$

where $r_i \sim N(0.5,1)$ is item feature relevant to objective 1, $c_{u,i}$ represents the personalized and contextualized relationship between two objectives, $L[.]$ is the logit model, and $\eta_{u,i} \sim N(0, 0.1)$ is the error term. We simulate for 100 users and 100 items, resulting in 10,000 records in total, and then provide recommendations using two versions of Shared-Bottom model (Ruder 2017), with and without knowing $c_{u,i}$, and compare their performance with the theoretical Pareto optimality that we can possibly achieve from this dataset. As we show in Figure 2, by taking into account the tradeoff between the two objectives, we will be able to significantly improve the Pareto frontier of recommendations across the entire dataset, and take it closer to the Pareto optimality.

Motivated by this example, in this paper, we propose a novel method to capture the tradeoff between different objectives and to approach the Pareto optimality. Our method is built upon a state-of-the-art multi-task learning method HyperNetwork (Ma et al. 2020; Hoang et al. 2023), where we enhance it through an attention mechanism and connect it to the reinforcement learning technique, which is described in the next section. By doing so, we will empirically outperform existing solutions in the literature, as well as achieve nice theoretical properties.

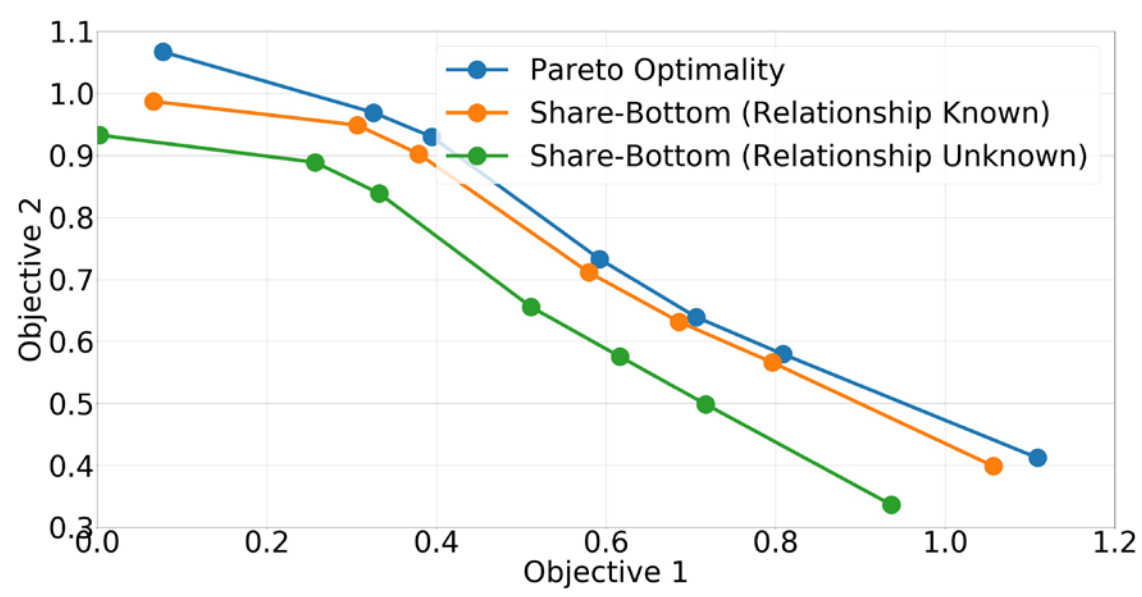

*Figure 2: Simulation Example for Illustrating the Importance of Personalized & Contextualized Information in Multi-Objective Recommendations*

### 3.4 Reinforcement Learning

Evidence presented in Sections 3.2 and 3.3 motivates the importance of modeling personalized and contextualized information. To achieve this goal, we select the reinforcement learning method (RL) (Sutton & Barto 2018) since it is especially useful for capturing such information and updating recommendations accordingly (Zheng et al. 2018). In addition, RL can also optimize the long-term recommendation performance, which is crucial for industrial platforms to maintain competitiveness (Zou et al. 2019; Chen et al. 2019; Afsar et al. 2022) since many of them suffer from the degradation of performance in the long run.

Note that our setting is significantly different from those studied in the literature (Liebman et al. 2019), as we aim to optimize *multiple* objectives, rather than only one objective. Following the RL terminology, we formulate our task for consumer $u$ at time $t$ with these notations:

- **State $\boldsymbol{s_{u,t}}$**: the personalized information of consumer $u$ for the current recommendation.
- **Action $\boldsymbol{a_{u,t}}$**: a set of objective weights $a_{u,t} = \{\alpha_{u,t,1}, \alpha_{u,t,2} \cdots, \alpha_{u,t,n}\}$ determined by $s_{u,t}$.
- **Context $\boldsymbol{C_{u,t}}$:** relevant contextual information that affects selection of optimal actions.
- **Reward $\boldsymbol{R_{s_{u,t},C_{u,t}}(a_{u,t})}$**: the benefit of taking action $a_{u,t}$, with state $s_{u,t}$ and context $C_{u,t}$.

The goal is to select the optimal action $a_{u,t}$ to maximize $R_{s_{u,t},C_{u,t}}(a_{u,t})$. We propose a *Deep Pareto Reinforcement Learning* (DeepPRL) model that *simultaneously* achieves significant performance improvements across multiple objectives, and we will now present its details.

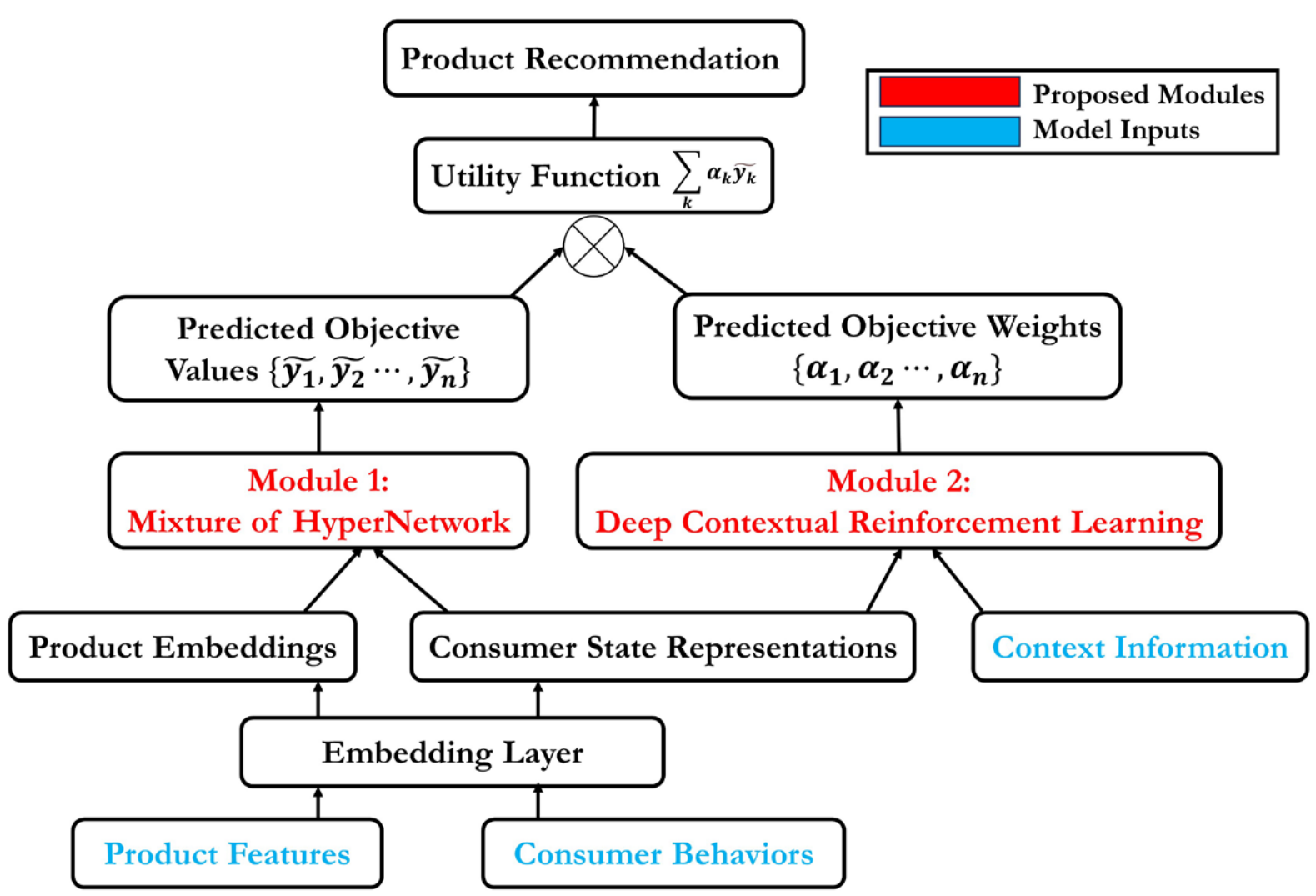

*Figure 3: Overview of Our Deep Pareto Reinforcement Learning Model*

# 4 Deep Pareto Reinforcement Learning Model

## 4.1 Overview of Our Proposed Model

We present the overall schema of DeepPRL in Figure 3. It takes product features, consumer behaviors, and contextual information (shown in blue) as inputs, and produces the output of multi-objective recommendations (shown at the top). The first step is to generate product embeddings and consumer state representations from the embedding layer (shown at the bottom). Then, these embeddings are fed into the "Mixture of Hypernetwork" module to predict the objective values $\{\widetilde{y_1}, \widetilde{y_2} \cdots, \widetilde{y_n}\}$ with $n$ hypernetworks and a mixture attention layer. We will discuss the details of this module in Section 4.2. Additionally, we also feed consumer state representations and the context information into our "Deep Contextual Reinforcement Learning" module to produce the set of objective weights $\{\alpha_1, \alpha_2 \cdots, \alpha_n\}$. We will describe the specifics in Section 4.3. In the final layer of Figure 3, we construct the aggregated utility function based on module outputs as $\sum_k \alpha_k \widetilde{y_k}$, and then provide recommendations with the highest utility values.

### 4.2 Mixture of HyperNetwork

The "Mixture of Hypernetwork" module is designed to tackle the challenge that dynamic objective relationships have not been properly modeled. It predicts the objective values with HyperNetwork (Ha et al. 2016), which operates by using one neural network to generate the weights for a series of networks. It thus enables a "soft" knowledge transfer between different learning tasks, so that when identifying the latent information of each objective, we also take into account other objectives by sharing the same neural network for generating the network parameters. This type of "soft" knowledge transfer was more effective than "hard" knowledge transfer (such as Shared-Bottom in Figure 4(a)), both theoretically and empirically, in terms of learning task relationships (Mahabadi et al. 2021; Shamsian et al. 2021; Beck et al. 2023).

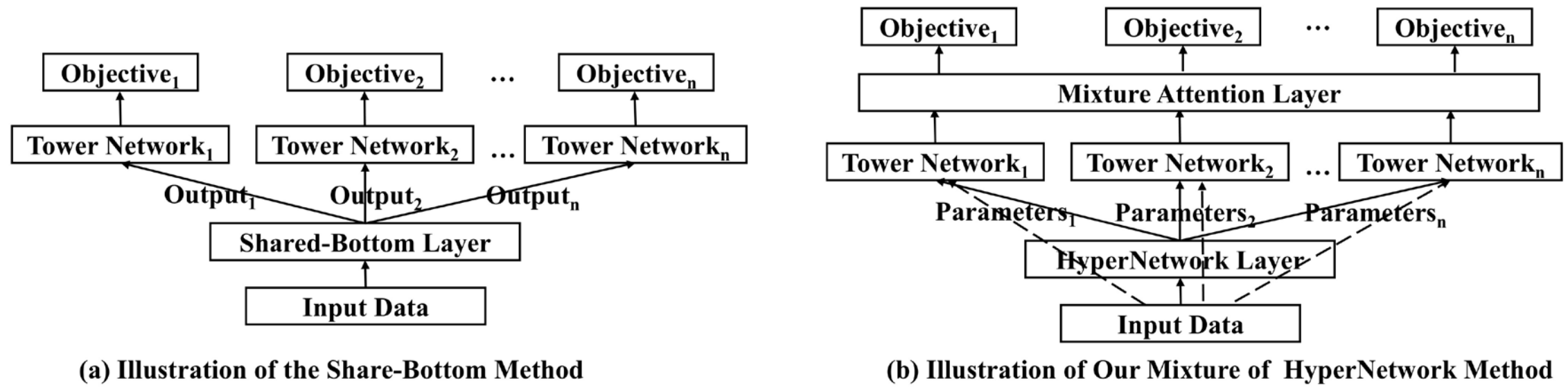

*Figure 4: Comparison of Share-Bottom and Our Mixture of HyperNetwork Methods*

To this end, we construct a HyperNetwork Layer, as shown in Figure 4(b), to capture the latent information from product and consumer embeddings. The output of the hypernetwork is a series of parameters and weights for constructing the corresponding tower networks, which will produce hidden embedding vectors containing objective-specific latent information. These vectors are then fed into a mixture attention network to jointly predict the objective values $\{\widetilde{y_1}, \widetilde{y_2} \cdots, \widetilde{y_n}\}$ (see Figure 3), as we assign different attention values to different objectives to model the heterogeneous relationships between them. The overall architecture of Mixture of HyperNetwork is presented in Figure 4(b) and works as follows. For product $i$, its objective values $\widetilde{y_{u,\iota,t}} = \{\widetilde{y_{u,\iota,t,1}}, \widetilde{y_{u,\iota,t,2}}, \dots, \widetilde{y_{u,\iota,t,n}}\}$ for consumer $u$ at time $t$ are predicted as:

$$\widetilde{y_{u,\iota,t,m}} = g_m\left(f_1([s_{u,t}; s_i]), f_2([s_{u,t}; s_i]), \dots, f_n([s_{u,t}; s_i])\right) \quad (1)$$

where $f_k(\cdot)$ is the $k$-th tower network (with parameters determined by the hypernetwork) that captures the latent information for the $k$-th objective, $[s_{u,t}; s_i]$ represents the concatenation of consumer state $s_{u,t}$ and product embedding $s_i$ that we obtained from the embedding layer in Figure 3, and $g_m(\cdot)$ is the mixture attention network that calibrates the outputs of hypernetworks for predicting the $m$-th objective. The attention value $e_{ml}$ for the $m$-th objective towards the $l$-th hypernetwork is computed following the self-attention mechanism (Shaw et al. 2018):

$$e_{ml} = \frac{\exp\left(d(W_m, W_l)\right)}{\sum_{j=1}^{n} \exp\left(d\left(W_m, W_j\right)\right)}$$

where $W_l$ is the output of the $l$-th hypernetwork and $d(.,.)$ is the Euclidean distance metric. The module is optimized by minimizing the difference between $\widetilde{y_{u,\iota,t}}$ and the ground truth $y_{u,i,t}$:

$$L(prediction) = \sum_{u,i,t} \sum_{k=1}^{n} \left|\left|\widetilde{y_{u,\iota,t,k}} - y_{u,i,t,k}\right|\right| \quad (2)$$

Specific configurations, such as the size and the structure of each neural network, will be discussed in Section 5. The novelty of our Mixture of HyperNetwork module lies in that we construct multiple neural networks to capture the objective-specific latent information respectively (rather than directly predicting the objective values), and construct a mixture-attention layer on top of these tower networks to calibrate the relative importance of each piece of latent information when predicting the values of recommendation objectives. As a result of these innovations, our model is more effective in sharing the latent information between different neural networks, and in learning the complex relationships between multiple objectives for the purpose of providing Pareto-optimal recommendations, as we demonstrate in the paper.

**4.3 Deep Contextual Reinforcement Learning**

To determine the optimal balance between different objectives, it is crucial to take into account personalized and contextualized consumer preferences. In this section, we present a *Deep Contextual Reinforcement Learning* method (see Module 2 in Figure 3) to determine the weights of multiple objectives. Our method is motivated by Reinforcement Learning (RL), which can effectively model personalized preferences in recommender systems (Afsar et al. 2022), as discussed in Section 3.4. However, our method also incorporates contextualized information into RL to address contextual fluctuations of consumer preference towards different objectives.

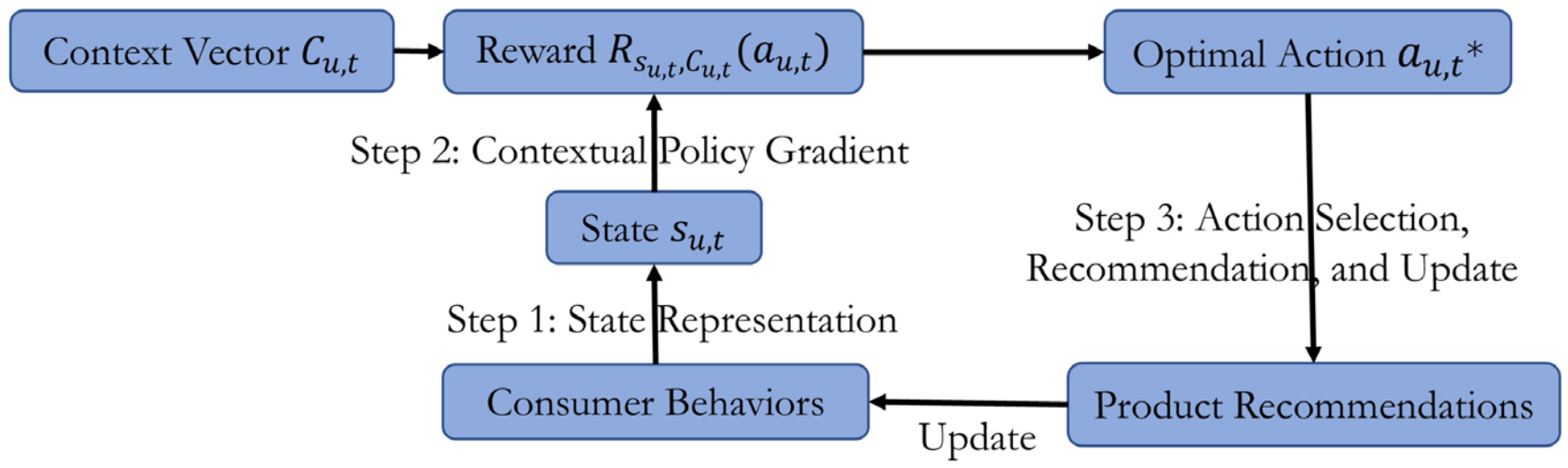

*Figure 5: Overview of the Deep Contextual Reinforcement Learning Module*

Specifically, our module consists of a three-stage learning process, as shown in Figure 5, and works as follows. In Step 1 (*State Representation*), we construct the representation of the consumer state $s_{u,t}$ as a latent embedding for effectively modeling personalized preference. While there are many models that can achieve this task, such as RNN, LSTM or Transformer (Sun et al. 2019), we present a *Self-Attentive Time-LSTM* method in this paper that provides the following modeling advantages: it (1) effectively models the temporal dynamics of consumer behaviors through sequential neural network (Wang et al. 2019); (2) captures time information between consumers' consecutive transactions to better understand the sequential decision-making process (Zhu et al. 2017); and (3) utilizes the Self-Attention mechanism (Shaw et al. 2018) to assign different weights to transactional records to account for their heterogeneous impact on consumer decisions. As a result, it captures the essence of personalized information more effectively than other alternatives, including BERT/Transformer (Vaswani et al. 2017). This is the case, as the latter performs well when modeling bidirectional contextual relations across words in a sentence, while the user purchase sequence is unidirectional and temporal. In addition, the multi-head attention mechanism in Transformer does not fit well with the sequential recommendation task, while self-attention usually works better (Kang & McAuley 2018).

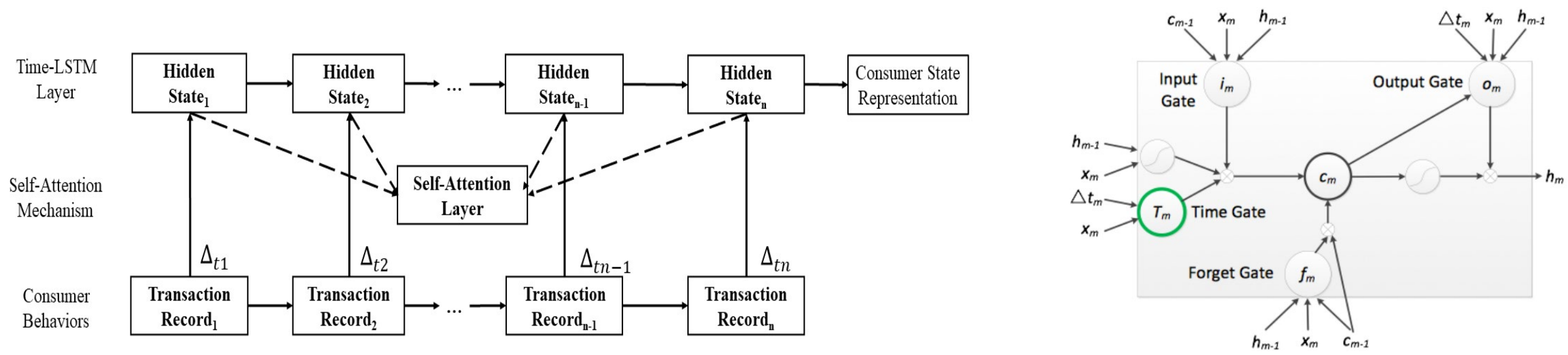

*(a) Overview of Our Self-Attention Time-LSTM Model (b) Time-LSTM Adapted from (Zhu et al. 2017)*

*Figure 6: Illustration of the Self-Attention Time-LSTM Model for Consumer State Representation*

Our Self-Attention Time-LSTM model is illustrated in Figure 6(a) and works as follows. We denote the transaction record of consumer u as $H_u = \{(TR_1, \Delta_{t1}), (TR_2, \Delta_{t2}) \cdots, (TR_n, \Delta_{tn})\}$, where $TR_k$ represents the $k$-th product, and $\Delta_{tk}$ stands for the time interval before the purchase of $TR_k$. The LSTM network consists of three components that control the information flow: the input gate $i_k$, forget gate $f_k$, and output gate $o_k$. Input $x_k = (TR_k, \Delta_{tk}, C_k)$, where $C_k$ is the context vector, $h_k$ represents the output hidden vector, and $c_k$ is the cell activation vector. We denote the activation function for each gate/vector as $\sigma_i$, $\sigma_f$, $\sigma_o$, $\sigma_c$, $\sigma_h$, the weight parameters as $W_{xi}$, $W_{xf}$, $W_{xo}$, $W_{hi}$, $W_{hf}$, $W_{ho}$, and the bias terms as $b_i$, $b_f$, $b_o$ respectively. To incorporate the time interval information into training, we add a time gate $T_k$, and update the hidden vector as:

$$i_k = \sigma_i(\widetilde{x_k} W_{xi} + h_{k-1} W_{hi} + b_i)$$

$$f_k = \sigma_i\left(\widetilde{x_k} W_{xf} + h_{k-1} W_{hf} + b_f\right)$$

$$o_k = \sigma_i(\widetilde{x_k} W_{xo} + h_{k-1} W_{ho} + b_o)$$

$$T_k = \sigma_t(\widetilde{x_k} W_{xt} + \sigma_\Delta(\Delta_k W_{ht}) + b_t)$$

$$c_k = f_k \otimes c_{k-1} + i_k \otimes T_k \otimes \sigma_c(\widetilde{x_k} W_{xc} + h_{k-1} W_{hc} + b_c)$$

$$h_k = o_k \otimes \sigma_h(c_k)$$

Note that the network inputs $\widetilde{x_k}$ are calibrated through the self-attention mechanism as the weighted sum of the linearly transformed inputs $x_k$, before being fed into the Time-LSTM:

$$\widetilde{x_k} = \sum_{i=1}^{k} e_{ki} \times W_{attention} x_i$$

$$e_{ki} = \frac{\exp\big(d(x_k, x_i)\big)}{\sum_{j=1}^{n} \exp\Big(d\big(x_k, x_j\big)\Big)}$$

where $W_{attention}$ stands for self-attention parameters, $e_{ki}$ represents the weight coefficients, and $d(x_k, x_i)$ is the similarity between inputs $x_t$ and $x_i$. By iteratively computing hidden states $h_k$ in Self-Attentive Time-LSTM, we obtain the hidden state $h_t$ at time $t$ based on the aforementioned equations, which constitutes the consumer state $s_{u,t} = h_t$, as shown in Figures 5, 6(a) and 6(b).

The novelty of our Self-Attentive Time-LSTM method presented in the paper lies in that we introduce an additional attention layer on top of Time-LSTM to capture heterogeneity in the sequence of the previously purchased products. To do that, we substitute all $x_k$ with $\widetilde{x_k}$, which is the calibrated input vector generated by the self-attention layer. In addition, rather than using Time-LSTM to produce recommendations directly (Zhu et al. 2017), we use Self-Attentive Time-LSTM to construct state representations for the subsequent reinforcement learning process, which enables us to update our policy in real-time based on the consumer feedback. As a result of these innovations, our method generates significantly better performance over alternatives.

Based on the consumer state described in Step 1, we next present our proposed *Contextual Policy Gradient* method in Step 2 (presented in Figure 5) to quantify the benefit of a specific action $a_{u,t}$. In the literature, such benefits are usually determined through Q-learning (Watkins & Dayan 1992; Mnih et al. 2015), where we select the optimal action that optimizes the Q-value.

However, this method only works well for discrete actions within a bounded action space, while in our setting, each action is a set of objective weights that can take arbitrary values, resulting in a large continuous action space. In addition, existing methods do not consider the contextual information in reinforcement learning. To this end, our Contextual Policy Gradient method works by developing two sets of neural networks: the actor network $\mu(s_{u,t}, C_{u,t} | \theta^{\mu})$ that selects a certain action $a$ based on the state $s_{u,t}$ and context $C_{u,t}$; and the critic network $Q(s_{u,t}, a_{u,t}, C_{u,t} | \theta^{Q})$ that generates the Q-value of the selected action $a_{u,t}$ based on state $s_{u,t}$ and context $C_{u,t}$. $\theta^{\mu}$ and $\theta^{Q}$ represent the parameters for the actor and critic networks respectively, which are optimized following the policy gradient technique (Lillicrap et al. 2016), where a policy $\pi$ is defined as the probability distribution of actions given a state, i.e., $\pi_{\theta}(a_{u,t} | s_{u,t})$. Since we cannot optimize the action $a$ directly, we will optimize its parameterized distribution instead to maximize the expected return $E\left[Q\left(s_{u,t}, C_{u,t}, \mu(s_{u,t}, C_{u,t} | \theta^{\mu})\right)\right]$. According to (Sutton & Barto 2018), the derivative of the expected return can be calculated by applying the chain rule:

$$\nabla_{\theta^{\mu}} J(\theta^{\mu}) \approx \frac{1}{N} \sum_{t} \nabla_{a} Q\left(s_{u,t}, a_{u,t} = \mu(s_{u,t}, C_{u,t} | \theta^{\mu})\right) \times \nabla_{\theta^{\mu}} Q\left(s_{u,t}, \mu(s_{u,t}, C_{u,t} | \theta^{\mu})\right) \qquad (4)$$

The critic network is optimized by minimizing the prediction error of the expected return, i.e.,

$$L(\theta^{Q}) = \frac{1}{N} \sum_{t} \left| Q(s_{u,t}, C_{u,t}, a | \theta^{Q}, t) - Q(s_{u,t}, C_{u,t}, a_{u,t}) \right|^{2} \qquad (5)$$

In addition, we update the parameters $\theta^{\mu}$ and $\theta^{Q}$ based on equations (4) and (5) following a soft updating mechanism (Van et al. 2016) to improve the learning stability:

$$\theta^{\mu} = \tau \theta^{\mu} + (1 - \tau) \theta^{\mu'}, \quad \theta^{Q} = \tau \theta^{Q} + (1 - \tau) \theta^{Q'} \qquad (6)$$

where $\theta^{\mu}$ and $\theta^{Q}$ are primary parameters that will be updated every iteration; $\theta^{\mu'}$ and $\theta^{Q'}$ are target parameters that will be fixed for some iterations and synchronized with $\theta^{\mu}$ and $\theta^{Q}$ periodically; and $\tau$ is the temperature that controls the update rate ($\tau \ll 1$). We use the primary networks to select the action and employ the target networks to estimate the Q-values.

The novelty of our Contextual Gradient Method lies in that we incorporate the contextual information when determining the most suitable action based on the consumer state, and conduct the reinforcement learning through the policy gradient technique to tackle the problem of large action space, enabling us to produce recommendations in an effective and efficient manner.

Finally, in Step 3 (*Recommendation and Update*) presented in Figure 5, we produce recommendations by selecting those products with the highest utility values $\sum_k \alpha_k \widetilde{y_k}$, where $\{\widetilde{y_1}, \widetilde{y_2} \cdots, \widetilde{y_n}\}$ are predicted by "Mixture of HyperNetwork", and $\{\alpha_1, \alpha_2 \cdots, \alpha_n\}$ are produced by "Deep Contextual Reinforcement Learning", as shown in Figure 3. Based on the feedback of consumer $u$, we update the state $s_{u,t}$, and context $C_{u,t}$ to produce subsequent recommendations.

### 4.4 Theoretical Properties of Our DeepPRL Model

In this section, we will present the theoretical properties of our method, which are related to the existence of Pareto optimality in our problem and the performance level of our model.

**Proposition 1 (Existence of Pareto Optimality)** For any multi-objective recommendation task formulated with the utility $\sum_k \alpha_{u,i,k} y_{u,i,k}$ involving *n* objectives $\{y_{u,i,1}, y_{u,i,2}, \dots, y_{u,i,n}\}$ for product $i$ and consumer $u$ at any time, there exists a Pareto optimal solution $\tilde{\alpha} = \{\widetilde{\alpha_{u,\iota,1}}, \widetilde{\alpha_{u,\iota,2}}, \cdots, \widetilde{\alpha_{u,\iota,n}}\}$ such that the aggregated utility $\sum_u \sum_i \sum_k \widetilde{\alpha_{u,\iota,k}} y_{i,k}$ is maximized.

**Proof.** According to the definition of Pareto optimality in multi-objective optimization (Navon et al. 2021), it is equivalent to showing that there exists $\tilde{\alpha}$ such that $\sum_u \sum_i \sum_k \widetilde{\alpha_{u,\iota,k}} \nabla L_k(u, i, y_{i,k}) =$ 0, where $L_k(u, i, y_{i,k})$ is the loss function for optimizing the $k$-objective when recommending product $i$ to consumer $u$. Since the utility is modeled as a linear function, the loss function (Equation (2) in Section 4.2) becomes continuously differentiable, and its Jacobian becomes invertible. Then, based on the implicit function theorem (Krantz & Parks 2002), there exists a neighborhood $U$ such that for any $\alpha \in U$, $\sum_u \sum_i \sum_k \alpha \nabla L_k(u, i, y_{i,k}) = 0$, Q.E.D.

**Proposition 2 (Performance Bound)** For any multi-objective recommendation task formulated in Section 3.1, the difference between the loss (empirical risk) $\varepsilon$ achieved by our DeepPRL model in Section 4, and the minimum loss $\varepsilon^*$ achieved by the Pareto optimal is bounded.

**Proof.** The main idea is to decompose the differences $|\varepsilon - \varepsilon^*|$ into multiple disjoint components, and obtain the upper bound for each one respectively. Specifically, for $n$ different objectives, we denote the predictive function for each objective as $f_1, f_2, \dots, f_n$, which are all associated with our DeepPRL model. We also denote the loss function as $l(.)$, and the consumer state representation as $h$. Then for input $X$ and output $Y$, we can express the two empirical risk terms as follows:

$$\varepsilon(h, f_1, f_2, \dots, f_n) = \frac{1}{n}\sum_{i=1}^{n} E_{(X,Y)\sim\mu_i} l(f_i(h(X)), Y)$$

$$\varepsilon^*(h, f_1, f_2, \dots, f_n) = \min_{h\in H, (f_1, f_2, \dots, f_n)\in F} \varepsilon(h, f_1, f_2, \dots, f_n)$$

where $\mu_i$ is the measure for the $i$-th objective, $H$ and $F$ are the functional spaces for $h$ and $f_i$. For the case of Pareto optimality, we denote the optimal predictive function as $f_1^*, f_2^*, \dots, f_n^*$, and:

$$\varepsilon - \varepsilon^* = \left(\varepsilon(h, f_1, f_2, \dots, f_n) - \frac{1}{nT}\sum_{it} l\big(f_i\big(h(X_{it})\big), Y_{it}\big)\right) +$$

$$\left(\frac{1}{nT}\sum_{it} l\big(f_i\big(h(X_{it})\big), Y_{it}\big) - \frac{1}{nT}\sum_{it} l\big(f_i^*\big(h(X_{it})\big), Y_{it}\big)\right) +$$

$$(\frac{1}{nT}\sum_{it} l\big(f_i^*\big(h(X_{it})\big), Y_{it}\big) - \frac{1}{n}\sum_{i=1}^{n} E_{(X,Y)\sim\mu_i} l(f_i^*(h(X)), Y))$$

The first term is bounded by $\sup_{h\in H,(f_1,f_2,\dots,f_n)\in F} \varepsilon(h, f_1, f_2, \dots, f_n) - \frac{1}{nT}\sum_{it} l\big(f_i\big(h(X_{it})\big), Y_{it}\big)$, which, according to the Benefits of Multi-Task Representation Learning (Theorem 13 in (Maurer et al. 2016)), will be further bounded by a constant that only involves $|\sup_{h\in H} h(X)|$. The second term is bounded, since the predictive function $f_i$ is learned using Stochastic Gradient Descent to approximate the optimal function $f_i^*$, and the Convergence Theorem (Bottou 1998) states that the difference between them will be bounded by a constant that only involves the input $X$ and the learning rate. Finally, the third term only involves $nT$ random variables $l\big(f_i^*\big(h(X_{it})\big), Y_{it}\big)$ sampled from a uniform distribution. Since these variables refer to the loss of the optimal predictive function $f_i^*(.)$, which only involves random noise when the predictive function is sufficiently trained. Therefore, these variables are independent, and according to Hoeffding's Inequality (Hoeffding 1994), the difference will be bounded with probability $1 - \frac{\delta}{2}$ by $\sqrt{\ln\left(\frac{2}{\delta}\right)/(2Tn)}$. A union bound of these three components will complete the proof. Q.E.D.

In conclusion, these two propositions demonstrate that the Pareto optimality exists, and that DeepPRL will achieve a performance level where its difference from the Pareto optimality is bounded. Additionally, according to (Hoang et al. 2023), the hypernetwork technique enables us to converge to the Pareto front faster than other alternatives, and produce the best performance in terms of hypervolume maximization. Therefore, we will be able to explicitly specify the conditions under which multiple recommendation objectives will constitute Pareto optimality with our DeepPRL model: (1) each objective takes only numeric values between 0 and 1; and (2) the utility function should be a continuously differentiable function with an invertible Jacobian. Note that those two conditions are very mild and can be easily satisfied across a wide range of scenarios, since the former can be achieved through normalization of the objective values, while the latter can be satisfied with the selection of a weighted linear function, as we use in this paper.

### 4.5 Summary of Our Proposed DeepPRL Model

In contrast to the existing methods, our approach explicitly captures the dynamic objective relationships through hypernetworks and determines the right balance through contextual reinforcement learning, thus providing the following benefits. First, it enhances the level of personalization and contextualization in multi-objective recommendations by selecting the appropriate set of objective weights for *each* consumer under *each* contextual scenario, resulting

**Algorithm 1** Deep Pareto Reinforcement Learning (DeepPRL)

**Require:** dataset $D$, consumer features $e_u$, product features $e_i$, contextual features $C$, ground-truth objectives $\{y_1, y_2, ..., y_n\}$, ground-truth reward $r$

***Mixture of HyperNetwork***:

**for** consumer $u$, product $i$ at time $t \in \mathcal{D}$ **do**
    **Product Embedding Generation**: Generate the product embedding $W_i$ by feeding product features $e_i$ into the embedding layer
    **Consumer State Representation**: Generate the consumer state representation $s_{u,t}$ by feeding the consumer features $e_u$ at time $t$ into the Self-Attention Time-LSTM network described in Section 4.3
    **Objective Prediction and Model Update**: Generate predictions of multiple objectives $\{y_{u,\tilde{i},t,1}, y_{u,\tilde{i},t,2}, ..., y_{u,\tilde{i},t,n}\}$ following Equation (1), compute $\mathcal{L}_{\text{prediction}}$ as in Equation (2), and update the module parameters
**end for**

***Deep Contextual Reinforcement Learning***:

**for** consumer $u$, product $i$ at time $t \in \mathcal{D}$ **do**
    **Contextual Policy Gradient** Construct the actor network $\mu(s_{u,t}, C_{u,t}|\theta^{\mu})$ and the critic network $Q(s_{u,t}, C_{u,t}, a_{u,t}|\theta^{Q})$
    **Model Update** Generate the action $a_{u,t} = \{\alpha_{u,t,1}, \alpha_{u,t,2}, ..., \alpha_{u,t,n}\}$ through the actor network, observe the reward $r_{s_{u,t},C_{u,t}}(a_{u,t})$, and update the parameters $\theta^{\mu}$ and $\theta^{Q}$ based on $\nabla_{\theta^{\mu}} J(\theta^{\mu})$ and $L(\theta^{Q})$ in Equations (4) and (5) respectively
**end for**

Produce **objective value predictions** $\{y_{u,\tilde{i},t,1}, y_{u,\tilde{i},t,2}, ..., y_{u,\tilde{i},t,n}\}$ following the Mixture of HyperNetwork component

Produce **objective weights** $\{\alpha_{u,t,1}, \alpha_{u,t,2}, ..., \alpha_{u,t,n}\}$ following the Deep Contextual Reinforcement Learning component

**return** Recommend product to maximize the utility $\sum_{u,i,t} \sum_{k=1}^{n} \alpha_{u,i,k} \times y_{u,\tilde{i},t,k}$

in significant performance improvements. Second, it automatically and effectively identifies consumer preferences towards different objectives from archival records, without requiring any explicit consumer feedback through surveys, as was done in the literature (Swait et al. 2018). Third, by incorporating future rewards during the reinforcement learning process (rather than only immediate rewards), our model also optimizes the long-term (versus only the short-term) performance of multiple objective recommendations, which is particularly useful for many platforms relying on the business performance over the long run (Zou et al. 2019). Fourth, we mitigate the data sparsity and scalability problems by updating the weights in real-time through deep learning techniques, making our method even more suitable for large-scale industrial platforms. Finally, all these prior advantages enable us to reach a level of performance that significantly outperforms existing state-of-the-art solutions, which we will demonstrate in Sections 5 and 6. Our model is also general in the sense that it works for any number of objectives since action $a$ can represent an arbitrary number of weights, making it particularly useful in real-world applications with more than two objectives (Zheng & Wang 2022). The pseudocode of DeepPRL is presented in Algorithm 1 above.

| **Model** | **Objectives Relationships** | **Dynamic Weighting** | **Contextual Information** | **Pareto Optimization** | **Long Term Performance** |
|---|---|---|---|---|---|
| **DeepPRL (This Paper)** | √ | √ | √ | √ | √ |
| MMOE (Ma et al. 2018) | √ | √ | × | × | × |
| Shared Bottom (Ruder 2017) | √ | × | × | × | × |
| DIN (Zhou et al. 2018) | × | × | √ | × | × |
| BERT4Rec (Sun et al. 2019) | × | × | √ | × | × |
| HyperNetwork (Hoang et al. 2023) | × | √ | × | √ | × |
| DJ-Monte Carlo (Liebman et al. 2019) | × | × | √ | × | √ |

*Table 1: Comparisons with Selected State-of-the-Art Methods in the Literature*

To further illustrate the advantages of our method, we compare it with previously proposed approaches across the following five dimensions in Table 1: (1) **Objective Relationships**, whether the model explicitly models the relationships between different recommendation objectives; (2) **Dynamic Weighting**, whether the model uses a fixed vs. a dynamic set of objective weights for producing recommendations; (3) **Contextual Information**, whether the model incorporates the contextual information in recommendations; (4) **Pareto Optimization**, whether the model optimizes one single objective, or optimizes the Pareto frontier of multiple objectives; (5) **Long Term Performance**, whether the model optimizes not only the short-term but also its long-term performance. We would like to note that "dynamic weighting" methods may or may not explicitly consider the "objective relationships" in the literature (and vice versa). For example, in (Hoang et al. 2023), the objective weights are determined dynamically based on a set of neural networks and the sequence of the input data, while the objective relationships are not considered at all during the subsequent recommendation process. On the other hand, (Ruder 2017) explicitly models the relationships between different objectives in the bottom layer of the neural network; however, the objective weights are determined as a set of fixed values based on the output of the neural network, and the method does not allow dynamic weighting as a result. We can observe from Table 1 that our model incorporates all five important aspects of the multi-objective recommendation task, leading to significant performance improvements over other alternatives, which we will demonstrate through offline and online experiments next.

# 5 Offline Experiment

## 5.1 Datasets and Recommendation Objectives

To evaluate the performance of our proposed method, we apply it to the following four datasets collected from major tech companies that we summarized in Table 2, each with two targeted recommendation objectives. We will also study a recommendation task with three targeted objectives in our online experiment, which we will describe in detail in the next section.

| Datasets | Alibaba-Youku | Yelp | Spotify | Kuaishou |
|---|---|---|---|---|
| # of Data Records | 1,806,157 | 2,254,589 | 970,013 | 4,676,570 |
| # of Consumers | 46,143 | 76,564 | 294,469 | 1,411 |
| # of Products | 53,657 | 75,231 | 81,948 | 3,327 |
| Density Levels | 0.0729% | 0.0004% | 0.0040% | 99.62% |

*Table 2: Descriptive Statistics of Four Datasets for Offline Experiments*

(1) **Alibaba-Youku**, which is a major video streaming platform in China (Li et al. 2020), and provides multi-objective short video recommendations by optimizing the *Video View* (number of videos viewed per session) and the *Dwell Time* (watching time spent per session) objectives.

(2) **Yelp**, which is a popular local business review site (Luca 2016), and offers multi-objective restaurant recommendations based on the *Relevance* (whether the recommended restaurant falls into the same category with the last visited restaurant) and *Novelty* (whether the recommended restaurant falls into a category that is different from all previous visits) objectives.

(3) **Spotify**, which is a media streaming site (Mehrotra et al. 2019) that produces multi-objective music recommendations based on the *Listening Percentage* (the percentage of a song the consumer listened to) and *Saved* (whether the consumer saved a song or not) objectives.

(4) **Kuaishou**, which is one of the largest short-form video-streaming platforms in China. The unique property of the KuaiRec dataset (Gao et al. 2022) lies in that it contains an almost fully observed user-item interaction matrix (density level=99.6%), i.e., each user has viewed each video and then left feedback, which alleviates the missing reward problem that we typically encounter when evaluating reinforcement learning models in the offline settings (Afsar et al. 2022). Video recommendations are generated according to the *Consistency* (content similarity between the recommended video and the last watched video) and *Recency* (the inverse exponential of dwell time between the video was posted and recommended) objectives.

**Relationships Between Different Objectives** Note that the four pairs of objectives in these applications constitute very heterogeneous relationships, both in terms of the nature of relations and the Pearson correlation coefficients, as we summarized in Table 3. Since they represent very different applications and vastly heterogeneous objective relationships, the offline results would be able to demonstrate the robustness and generalizability of our model. We shared the codes at https://anonymous.4open.science/r/Multi_Objective-CF5F for reproducibility purposes.

| Datasets & Objectives | Pearson Coefficients | Nature of Relationships |
|---|---|---|
| Alibaba-Youku (Video View, Dwell Time) | -0.238 | usually conflict with each other, as focusing on VV leads the system to recommend shorter videos to encourage finishing watching each video quickly, which is sub-optimal for DT |
| Yelp (Relevance, Novelty) | -0.596 | contradictory to each other; while relevant restaurants usually fit well with consumers' tastes, it is also beneficial to provide novel restaurant choices that might fit consumers' culinary preferences even better. |
| Spotify (Listening%, Saved) | 0.369 | usually positively correlated, but might still part ways, as a newly released piece of music generally has a higher listening percentage due to the novelty factor, but the saved metric might not necessarily be high. |
| Kuaishou (Consistency, Recency) | -0.024 | not correlated (independent) with each other, since these two objectives are measuring two orthogonal dimensions of consumer experience |

*Table 3: Summary of the Relationships Between Different Objectives in Offline Experiments*

**Business Relevance of the Selected Objectives** It is worth noting that all objectives that we select are of significant business relevance to the associated platform. Specifically, according to the written testimony provided by the engineering lead of Alibaba-Youku, *Video View* and *Dwell Time* are the most relevant objectives for generating revenues out of the streaming platform; both the *Relevance* and *Novelty* objectives are shown to be directly related to consumer experience on Yelp (Dai et al. 2018); based on the personal communications that we have with the then-director of research of Spotify, both listening percentage and saved are with most business significance for the company; and finally, according to (Jin et al. 2024), *Consistency* and *Recency* are the two most important objectives for Kuaishou. Therefore, performance improvements observed in our offline experiments are meaningful from a business perspective and result in significant economic impact.

**Reward Formulation** Following the standard practices in the literature (Zou et al. 2019; Asfar et al. 2022), we use "user activeness" as the reward function in our offline experiments. More specifically, for Alibaba-Youku and Kuaishou, the reward is a numeric value provided directly to us by the companies, measuring how "active" the user is on a recommended video. For Yelp, the reward is a numeric value of the rating that the user left for the recommended restaurant, while for Spotify, the reward is a numeric value measuring user satisfaction with the recommended song. These rewards are all relevant to our multi-objective recommendation task, since they are typically hard to optimize in an end-to-end manner directly, such as through back-propagation. Therefore, a common practice is to break down these rewards as a combination of multiple recommendation objectives (Zheng & Wang 2022), with the assumption that a high rating or a click behavior can be attributed to a mix of multiple factors, such as relevance and novelty of the recommended content. Therefore, we decompose the task of improving the reward of recommendations into two steps: (1) obtain more accurate objective predictions; and (2) better understand the relative contribution (i.e., weights) of each objective towards the final reward. As a result, we will be able to improve the Pareto Frontier of recommendation objectives, better model the relationship between them and the final reward, and achieve significant improvements in the reward as a result.

**Handling Missing Rewards** Note that in the offline evaluations, we may face the challenge of missing rewards, since we can only observe the feedback for products that are recommended to the users. Therefore, we follow the standard practices (Gunawardana et al. 2022) and restrict the candidate set to the set of products that each consumer has interacted with before so that we will have the reward information. To further address this challenge, we have also conducted analyzes on the fully-observed KuaiRec dataset and the online experiment at Alibaba-Youku. In both cases, we have the full reward information (no missing values) to ensure a comprehensive evaluation.

### 5.2 Baseline Models

In our experiments, we compare our proposed method with the following three groups of eight state-of-the-art baseline recommendation models, where the objective weights are all determined in a *static* manner through Bayesian Hyperparameter Optimization (Snoek et al. 2012):

(1) **Multi-Objective Recommendation Baselines**, including Multi-Objective Linear UCB (MO-LinCB) (Mehrotra et al. 2020), Multi-Objective Reinforcement Learning (MORL) (Abels et al. 2019), and Progressive Layered Extraction (PLE) (Tang et al. 2020).

(2) **Multi-Task Recommendation Baselines**, which include Multi-gate Mixture-of-Experts

(MMOE) (Ma et al. 2018), Mixture of Sequential Experts (MoSE) (Qin et al. 2020), and Shared-Bottom (Ruder 2017) models. We apply them to our application by jointly predicting and aggregating multiple objectives through a linear utility function with fixed weights.

(3) **Deep Reinforcement Learning Baselines**, which include REINFORCE (Chen et al. 2019) and DRN (Zheng et al. 2018) models. Although these two methods are single-objective, we have included them nevertheless to demonstrate that the advantages of our model do not come from the DRL technique alone; it requires a careful combination of the "Mixture of Hypernetworks" and "Deep Contextual Reinforcement Learning" modules that we propose in the paper.

We determine the optimal configurations for our model through Bayesian Hyperparameter Optimization (Snoek et al. 2012) and the hyperparameter sensitivity analysis presented in Section 5.7. As a result, Self-Attentive Time-LSTM includes two fully-connected layers with a batch size of 32. HyperNetwork is formulated as 3 fully-connected layers with the Sigmoid activation function. Contextual Policy Gradient is constructed as 3 fully-connected hidden layers with the Sigmoid activation function. Both the actor and critic networks are formulated as fully-connected networks. These models are optimized using the Stochastic Gradient Descent with a learning rate of 0.01. Finally, we select the discount rate $\gamma$ as 0.99, the temperature $\tau$ as 0.05, and number of hidden units as 64. Results are reported from time-stratified 5-fold cross-validation with 10 runs. For our model and all baselines. we use consumer demographics or characteristics as the personalized information, and the last 10 purchased products & timestamp as the contextual information, following standard industrial practices (Gomez-Uribe & Hunt 2015).

| Dataset | Alibaba-Youku | | Yelp | | Spotify | | Kuaishou | |
|---|---|---|---|---|---|---|---|---|
| Objective | VV | DT | Relevance | Novelty | Listening % | Saved | Consistency | Recency |
| **DeepPRL** <br><br> **(%Improved)** | **0.7486****** <br> **(0.0017)** <br> **+2.55%** | **0.9133****** <br> **(0.0021)** <br> **+2.27%** | **0.6814****** <br> **(0.0019)** <br> **+1.31%** | **0.2758****** <br> **(0.0008)** <br> **+23.46%** | **0.6033****** <br> **(0.0028)** <br> **+3.36%** | **0.3016****** <br> **(0.0019)** <br> **+3.75%** | **0.3459****** <br> **(0.0023)** <br> **+4.11%** | **0.8759****** <br> **(0.0029)** <br> **+1.80%** |
| MO-LinCB | 0.7208 | 0.8834 | 0.6680 | 0.2078 | 0.5799 | 0.2871 | 0.3179 | 0.8538 |
| MORL | 0.7233 | 0.8876 | 0.6698 | 0.2095 | 0.5814 | 0.2889 | 0.3192 | 0.8536 |
| PLE | 0.7279 | 0.8908 | <u>0.6725</u> | <u>0.2111</u> | 0.5823 | 0.2895 | 0.3263 | 0.8557 |
| MMOE | 0.7287 | 0.8917 | 0.6715 | 0.1990 | 0.5814 | 0.2889 | 0.3278 | 0.8577 |
| MOSE | <u>0.7295</u> | <u>0.8926</u> | 0.6723 | 0.2102 | <u>0.5830</u> | <u>0.2903</u> | <u>0.3317</u> | <u>0.8601</u> |
| Share-Bottom | 0.7231 | 0.8771 | 0.6677 | 0.1877 | 0.5814 | 0.2871 | 0.3256 | 0.8552 |
| REINFORCE | 0.7168 | 0.8798 | 0.6536 | 0.1911 | 0.5826 | 0.2865 | 0.3158 | 0.8512 |
| DRN | 0.7005 | 0.8744 | 0.6379 | 0.1897 | 0.5815 | 0.2871 | 0.3144 | 0.8502 |

*Table 4: Offline Experiments on Three Multi-Objective Recommendation Tasks. Improvements are shown over the second-baseline models (underlined). ***p<0.01.*

### 5.3 Experimental Results

The offline experimental results are shown in Table 4, where DeepPRL significantly outperforms all three groups of eight baseline models across all recommendation tasks. When compared with the second-best baselines, our model achieves significant improvements in *both* objectives *simultaneously* (e.g., 2.55% in VV and 2.27% in DT for Alibaba-Youku) across *all four* applications, as we *dynamically* select the most appropriate objective weights for *each* consumer. It also demonstrates the power of modeling multiple objectives simultaneously rather than focusing only on one objective. As these improvements are significant across all four pairs of objectives regardless of their heterogeneous relationships, we demonstrate the generalizability and flexibility of our model to *dynamically determine evolving balances between the objectives*, thus *significantly outperforming state-of-the-art baselines across all recommendation objectives*. We would also like to point out that the improvements that we achieved in the range of 1% - 23% (see Table 4) are considered to be very significant in the industry (Gomez-Uribe & Hunt 2015).

### 5.4 Pareto Frontier Analysis

In this section, we provide empirical evidence that our model dominates the Pareto frontiers of existing multi-task learning methods. To obtain the Pareto frontiers, we follow the established protocols (Ribeiro et al. 2014; Zheng & Wang 2022) and run each model multiple times with varying objective weights from 0 to 1, as we show in Figure 7(a-d). For our DeepPRL model, however, the objective weights are dynamically determined; therefore, we obtain its Pareto frontier by running it multiple times with different neural network initializations, following the practice in (Abels et al. 2019) for dynamic weighting models. We observe from Figure 7 that our model performs significantly better than all baselines in terms of *both* objectives across all four datasets and dominates their Pareto frontiers. In addition, our model obtains significantly higher convex hull volume compared to other alternatives, as illustrated in Table 5. In this respect, our model manages to improve the performance across multiple objectives *simultaneously* over the baselines, even for those conflicting ones, such as VV and DT in the Alibaba dataset. Therefore, we empirically demonstrate the advantages of our model in terms of Pareto optimization.

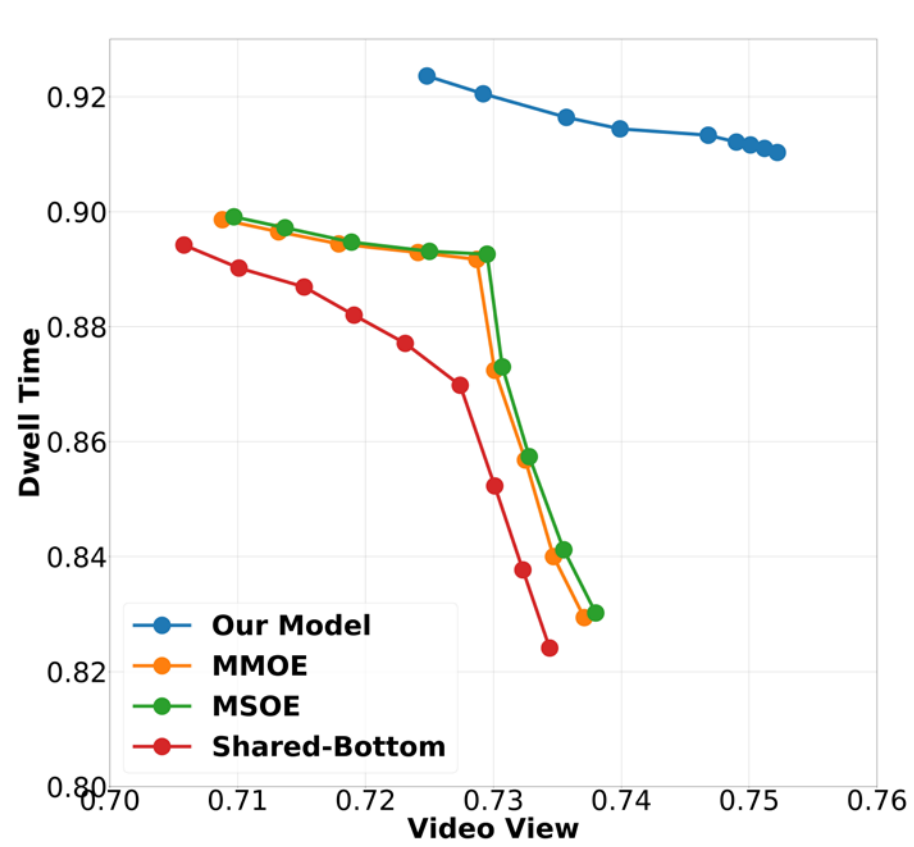

*(a) Alibaba-Youku dataset*

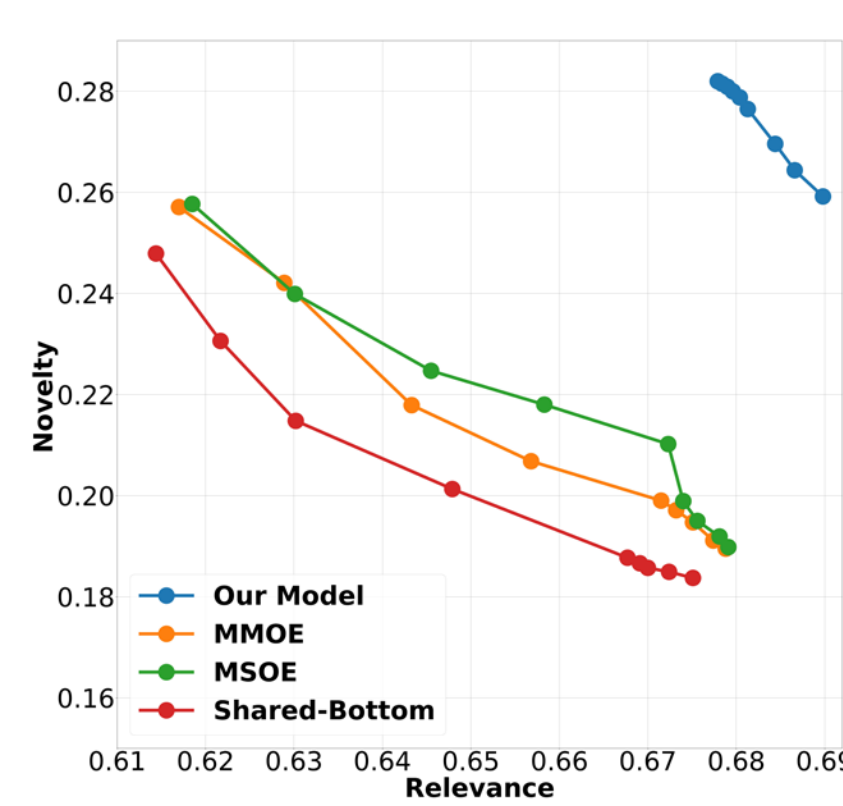

*(b) Yelp dataset*

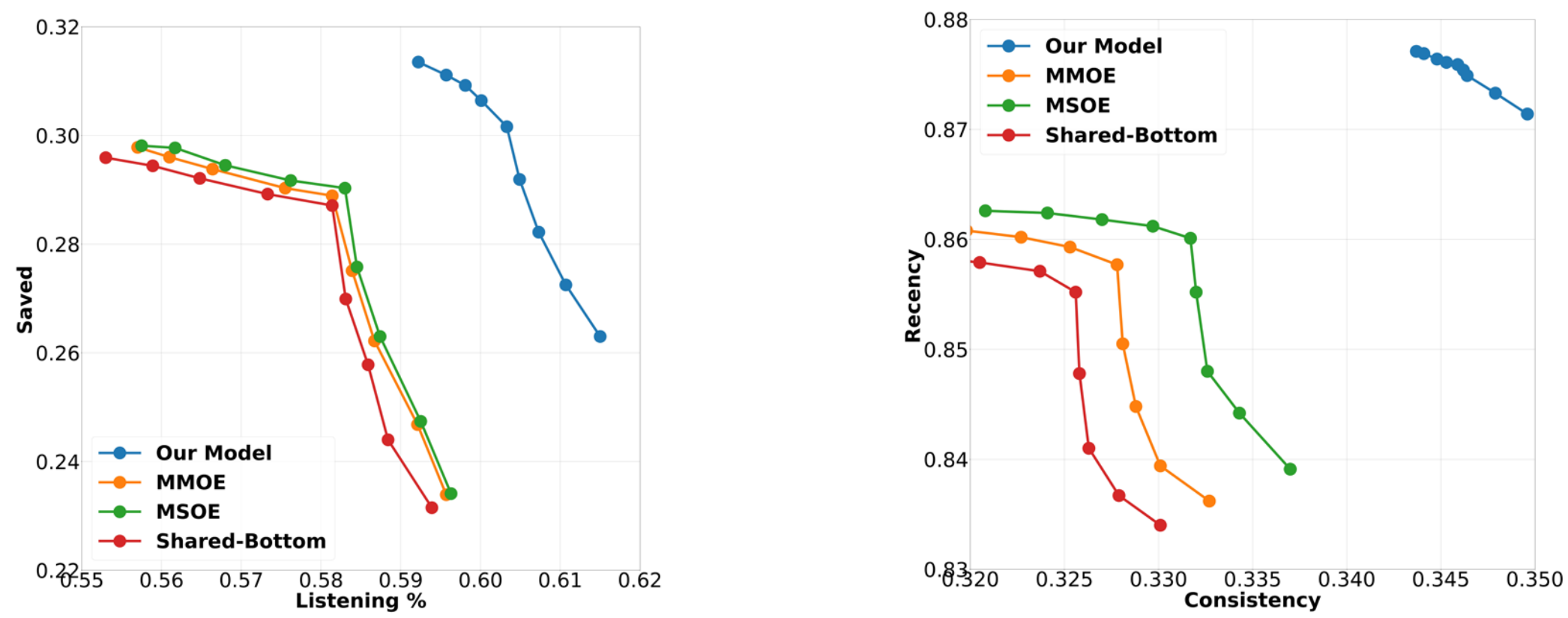

*(c) Spotify dataset* *(d) Kuaishou dataset*

*Figure 7: Pareto Frontier Plots with Different Sets of Importance Weights for Four Datasets*

| Dataset | Alibaba-Youku | Yelp | Spotify | Kuaishou |
|---|---|---|---|---|
| **DeepPRL** | **0.0011*** (0.0001)** | **0.0011*** (0.0001)** | **0.0015*** (0.0001)** | **0.0005*** (0.0001)** |
| MMOE | 0.0006 | 0.0005 | 0.0006 | 0.0001 |
| MOSE | 0.0006 | 0.0005 | 0.0007 | 0.0001 |
| Share-Bottom | 0.0005 | 0.0004 | 0.0005 | 0.0001 |

*Table 5: Comparisons of the Convex Hull Volume on Four Multi-Objective Recommendation Tasks. Improvements are shown over the second-baseline models (underlined). ***p<0.01.*

### 5.5. Analysis on the Role of Contextual Features and Dynamic Updating

To demonstrate how context and dynamic factors affect the outcome of our model, we conduct additional analysis in this section to compare with the following three variants: (1) **DeepPRL (w/o Dynamic Updating)**, where we learn fixed objective weights for each consumer instead using the Contextual Policy Gradient method described in Module 2 without dynamic updates; (2) **DeepPRL (w/o Contextualization)**, where we learn the dynamic objective weights using the standard Policy Gradient method without contextual features; and (3) **DeepPRL (w/o Dynamic Updating & Contextualization)**, where we learn fixed objective weights for each consumer using the standard Policy Gradient method without contextual features or dynamic updates.

Results in Table 6 show that both the dynamic and contextual factors are integral components in DeepPRL, without which the convex hull volume will decrease by 51.1% and 33.6% respectively, on average across four datasets. In addition, we pick a particular consumer (user_id "96212" in Alibaba-Youku) to see how both factors will alter the objective trade-off at the micro-level, where we repeat the Pareto Front Analysis introduced in the previous section to obtain the Pareto frontiers achieved by selected variants of our model. We can observe from Figure 8 that DeepPRL manages to reach a performance level almost identical to the pareto optimality, while without dynamic or contextual factors, all three variants only result in suboptimal performance, where the objective trade-offs all lie inside the convex hull of the pareto optimality.

| Dataset | Alibaba-Youku | Yelp | Spotify | Kuaishou |
|---|---|---|---|---|
| **DeepPRL** | **0.0011***<br>(0.0001)** | **0.0011***<br>(0.0001)** | **0.0015***<br>(0.0001)** | **0.0005***<br>(0.0001)** |
| DeepPRL (w/o Dynamic Updating) | 0.0006 | 0.0006 | 0.0007 | 0.0002 |
| DeepPRL (w/o Contextualization) | 0.0008 | 0.0008 | 0.0009 | 0.0003 |
| DeepPRL (w/o Dynamic Updating & Contextualization) | 0.0004 | 0.0004 | 0.0004 | 0.0001 |

*Table 6: Comparisons of the Convex Hull Volume on Different Variants of DeepPRL Model. ***p<0.01.*

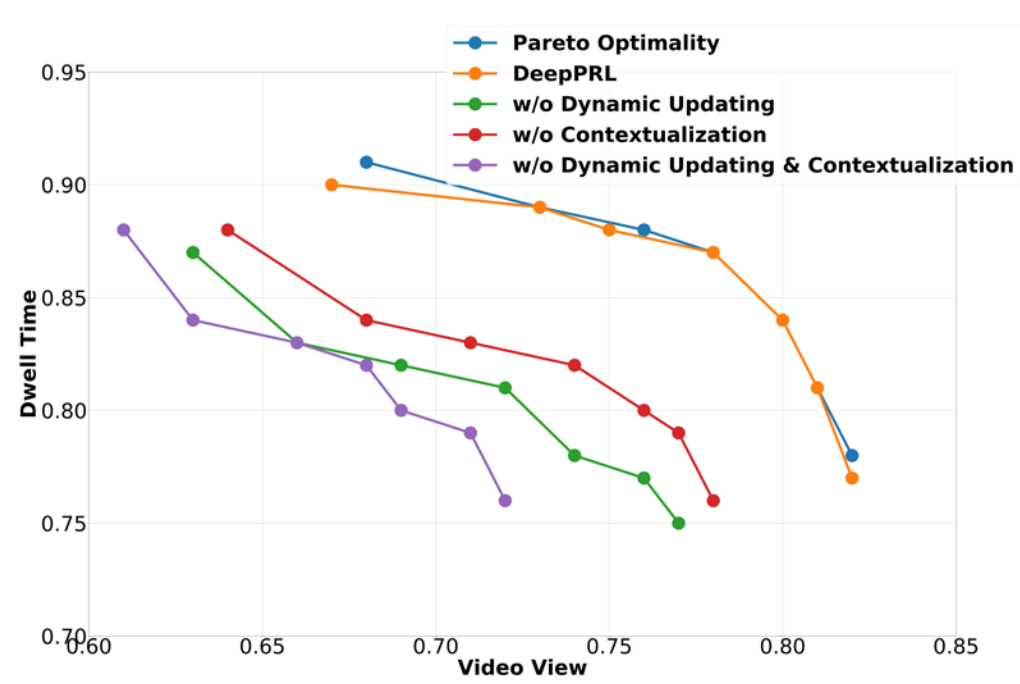

*Figure 8: Pareto Front Analysis on A Selected Consumer Picked From the Alibaba-Youku Dataset*

### 5.6 Ablation Study

Since our model consists of two modules, we conduct a series of ablation studies in this section to demonstrate the effectiveness of both modules and justify our design choices. We compare the performance of our model with the following variants summarized in Table 7, where we replace each component with an alternative design and keep the other intact:

| Model Variants | Descriptions |
| --- | --- |
| Variant 1 | Replace Self-Attentive Time-LSTM with Time-LSTM for state representations. |
| Variant 2 | Replace Self-Attentive Time-LSTM with Self-Attention LSTM for state representations. |
| Variant 3 | Replace Self-Attentive Time-LSTM with Transformer for state representations. |
| Variant 4 | Replace Mixture-of-HyperNetwork with Shared-Bottom to predict objective values. |
| Variant 5 | Replace Mixture-of-HyperNetwork with Cross Stitch to predict objective values. |
| Variant 6 | Replace Contextual Policy Gradient with DQN to produce objective weights. |
| Variant 7 | Replace Contextual Policy Gradient with REINFORCE to produce objective weights. |

*Table 7: Summary of Model Variants in the Ablation Study*

We report the results in Table 8, where we observe that our model significantly outperforms all seven variants. Specifically, improvements over Variants 1, 2, and 3 indicate the effectiveness of Self-Attention Time-LSTM for modeling consumer state representations; improvements over Variants 4 and 5 demonstrate the superiority of Mixture-of-HyperNetwork for predicting the values of multiple objectives; and improvements over Variants 6 and 7 show the effectiveness of Contextual Policy Gradient in maximizing the total return. To sum up, these ablation studies demonstrate that we need all the novel components described in this paper and their proper integration into a coherent model to produce superior recommendation performance.

| Dataset | Alibaba-Youku | | Yelp | | Spotify | | Kuaishou | |
| --- | --- | --- | --- | --- | --- | --- | --- | --- |
| Objective | VV | DT | Relevance | Novelty | Listening % | Saved | Consistency | Recency |
| **Our Model** | **0.7486*** (0.0017)** | **0.9133*** (0.0021)** | **0.6814*** (0.0019)** | **0.2758*** (0.0008)** | **0.6033*** (0.0028)** | **0.3016*** (0.0019)** | **0.3459*** (0.0023)** | **0.8759*** (0.0029)** |
| Variant 1 | 0.7304 | 0.8821 | 0.6661 | 0.2709 | 0.5888 | 0.2946 | 0.3273 | 0.8532 |
| Variant 2 | 0.7331 | 0.8761 | 0.6689 | 0.2718 | 0.5910 | 0.2941 | 0.3318 | 0.8566 |
| Variant 3 | 0.7344 | 0.8724 | 0.6708 | 0.2726 | 0.5933 | 0.2950 | 0.3342 | 0.8598 |
| Variant 4 | 0.7406 | 0.9025 | 0.6719 | 0.2724 | 0.5925 | 0.2957 | 0.3336 | 0.8583 |
| Variant 5 | 0.7399 | 0.8987 | 0.6717 | 0.2730 | 0.5941 | 0.2974 | 0.3350 | 0.8602 |
| Variant 6 | 0.7329 | 0.8813 | 0.6684 | 0.2695 | 0.5905 | 0.2944 | 0.3318 | 0.8583 |
| Variant 7 | 0.7357 | 0.8830 | 0.6693 | 0.2707 | 0.5925 | 0.2941 | 0.3328 | 0.8595 |

*Table 8: Ablation Studies on Three Multi-Objective Recommendation Tasks. ***p<0.01.*

### 5.7 Sparsity, Sensitivity, and Scalability Analysis

To further demonstrate the robustness of our framework under different experimental settings, we conduct additional analyses in this section. First, we randomly sample three subsets from the Spotify dataset with different sparsity levels ($Sparsity = (\text{\# of Transaction Records})/$ (# of Consumers × # of Products)) shown in Table 9, and we implement all models on them.

| Dataset | Spotify-1 | Spotify-2 | Spotify-3 |
|---|---|---|---|
| # Of Transaction Records | 970,013 | 91,523 | 11,921 |
| # Of Consumers | 294,469 | 12,373 | 865 |
| # Of Products | 81,948 | 55,903 | 9,588 |
| Sparsity | 0.004% | 0.0132% | 0.1437% |

*Table 9: Descriptive Statistics of Different Versions of the Spotify Dataset for Sparsity Analysis*

The results in Table 10 confirm that our model significantly outperforms baselines under all three datasets with different sparsity levels and different data sizes. We also observe that our model achieves the most improvements on the subset (Spotify-3) with the highest sparsity level and the lowest data size, demonstrating its effectiveness in handling the data-sparsity problem.

| Dataset | Spotify-1 | | Spotify-2 | | Spotify-3 | |
|---|---|---|---|---|---|---|
| Objective | Listening % | Saved | Listening % | Saved | Listening % | Saved |
| **Our Model** <br> **(%Improved)** | **0.6033****** <br> **(0.0028)** <br> **+3.36%** | **0.3016****** <br> **(0.0019)** <br> **+3.89%** | **0.6237****** <br> **(0.0044)** <br> **+5.23%** | **0.3110****** <br> **(0.0029)** <br> **+6.08%** | **0.6275****** <br> **(0.0061)** <br> **+5.75%** | **0.3124***** <br> **(0.0035)** <br> **+6.47%** |
| MO-LinCB | 0.5826 | 0.2886 | 0.5876 | 0.2901 | 0.5878 | 0.2907 |
| MORL | 0.5814 | 0.2901 | 0.5889 | 0.2905 | 0.5886 | 0.2914 |
| PLE | 0.5826 | 0.2898 | 0.5907 | 0.2917 | 0.5895 | 0.2918 |
| MMOE | 0.5814 | 0.2889 | 0.5889 | 0.2912 | 0.5892 | 0.2914 |
| MOSE | <u>0.5830</u> | <u>0.2903</u> | <u>0.5911</u> | <u>0.2921</u> | <u>0.5914</u> | <u>0.2922</u> |
| Share-Bottom | 0.5814 | 0.2871 | 0.5881 | 0.2886 | 0.5895 | 0.2889 |
| REINFORCE | 0.5826 | 0.2865 | 0.5901 | 0.2888 | 0.5906 | 0.2890 |
| DRN | 0.5815 | 0.2871 | 0.5910 | 0.2891 | 0.5913 | 0.2894 |

*Table 10: Sparsity Analysis on Spotify. Improvements over the second-baseline models (underlined). ***p<0.01.*

We further conduct the sensitivity analysis to study how different values of the discount rate $\gamma$ (in Equation 3), the temperature parameter $\tau$ (in Equation 6), and the number of hidden units affect the performance. As shown in Figure 9(a-c), the performance remains relatively steady and largely insensitive across different hyperparameter settings, as the fluctuations are only under 0.3% and small in their absolute values. The performance can be further improved when choosing the optimal values ($\gamma = 0.99$, $\tau$=0.95, Number of Hidden Units=64), as we did in Section 5.2.

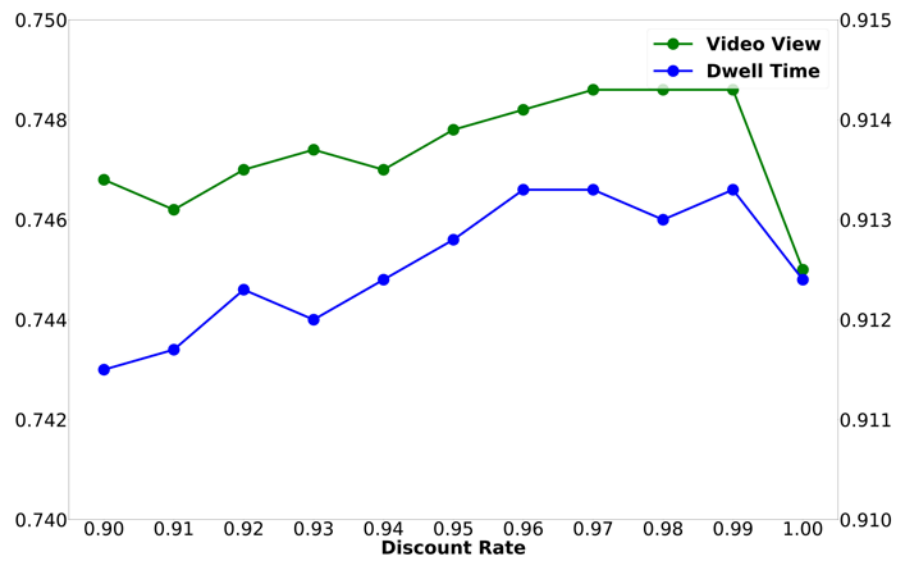

*(a) Discount Rate γ.*

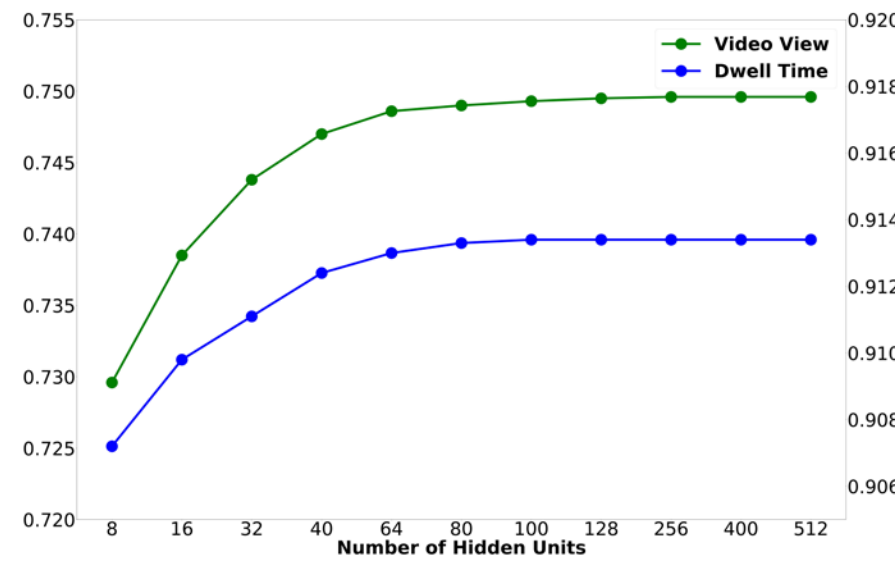

*(b) Number of Hidden Units.*

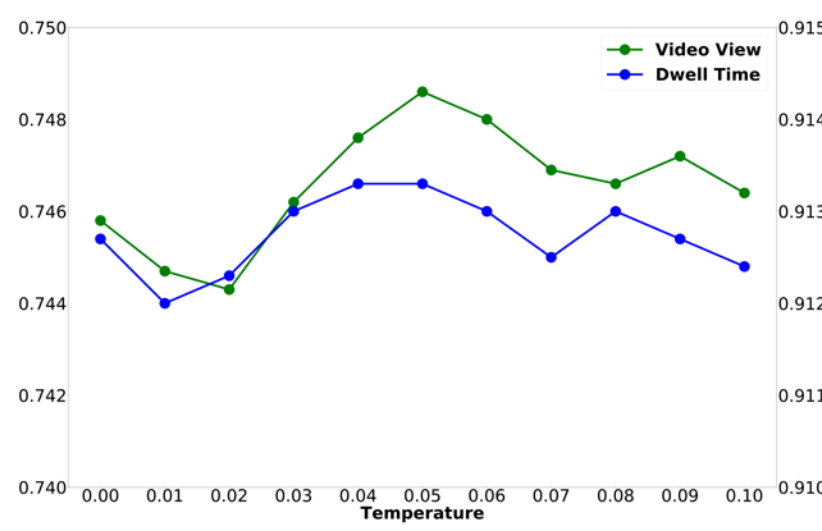

*(c) Temperature Parameter τ.*

*Figure 9: Sensitivity Study on Alibaba-Youku dataset with Different Hyperparameter Selection. Y-axis Shows the Average Value of the Objective*

Finally, to further demonstrate the practicality of our model, we conduct the scalability analysis where we train our model using a personal laptop on multiple randomly selected subsets of four offline datasets with different data sizes ranging from 100 to 1,000,000 records. We then record the training time for 100 epochs in Figure 10, where we observe that our model scales linearly in the number of data records, which comes from our use of deep learning techniques that are scalable and flexible in practice. In fact, according to (Yang & Amari 1998), the time complexity of our model with natural gradient descent is only $O(N)$, where $N$ is the data size.

To summarize, our proposed model provides scalable and effective recommendations and successfully deals with the data-sparsity problem, making it particularly useful in industrial settings, as we also demonstrate through the online controlled experiments in the next section.

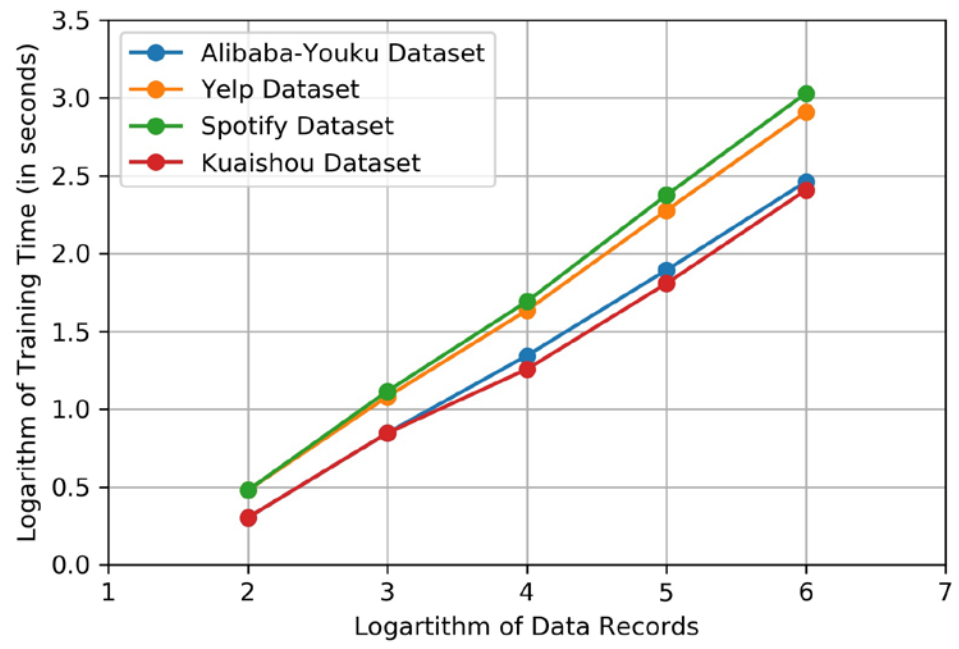

*Figure 10: Scalability Performance Analysis on the Four Offline Datasets.*

## 6 Long-Term Performance Analysis – A Simulation Study

An important advantage of our DeepPRL model lies in its capability of optimizing the long-term performance of multi-objective recommendations. However, there are very few archival datasets that include the counterfactual of how the subsequent recommendations will change accordingly if the consumer selects a different product. Therefore, we follow the paradigm advocated in (Zhang et al. 2020) and conduct a simulation study in this section to evaluate the long-term performance of our proposed model. Our simulation design (similar to Section 3.3) follows (Sun et al. 2019b), which is an important and popular paper in the CS literature that uses simulations to analyze long-term performance. Specifically, we generate the multi-objective objective matrices $O_{u,i,1}$ and $O_{u,i,2}$ using the discrete choice model in the item response theory (Embretson & Reise 2013) as:

$$o_{u,i,1} = L[r_i + \eta_{u,i}] \qquad o_{u,i,2} = 1 - c_{u,i} \times o_{u,i,1}$$

where $r_i \sim N(0.5,1)$ captures the features of item $i$ relevant to objective 1, $c_{u,i}$ represents the relationship between two objectives that is associated with personalized and contextual factors that are uniformly distributed, $L[.]$ is the logit model, and $\eta_{u,i} \sim N(0, 0.1)$ is the error term. We generate the matrices $O_{u,i,1}$ and $O_{u,i,2}$ with 100 consumers and 100 products, thus having 10,000 ratings in total. We first randomly select 10 products for each of the 100 users as their interaction records to start our initial training, and then select 80% of the objective matrices as the training data to optimize the parameters of our DeepPRL model. The remaining 20% will be used as the test data for evaluation purposes. In each iteration, we recommend the top-10 candidate products, and then stochastically simulate the consumers' response following (Sun et al. 2019b), and update the interaction history accordingly. In total, we simulate 50 iterations to make sure our model and baselines converge, and report performance as the average of 10 independent runs. We present the Pareto Frontier in Figure 11(a-d) for Iterations 1, 10, 25, and 50, respectively. We can observe from Figure 11 that DeepPRL consistently outperforms all baselines in terms of dominating their Pareto Frontiers, and that the performance improvements become larger when the number of iterations goes up (evidenced by the convex hull volume), indicating the superior performance of DeepPRL in the long-term due to the benefits of Deep Contextual Reinforcement Learning in our model.

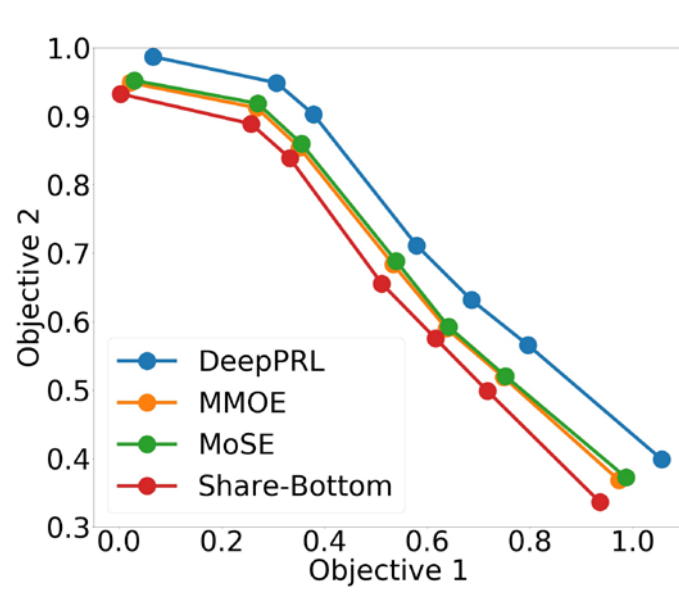

*(a) Iteration 1: +2.58%*

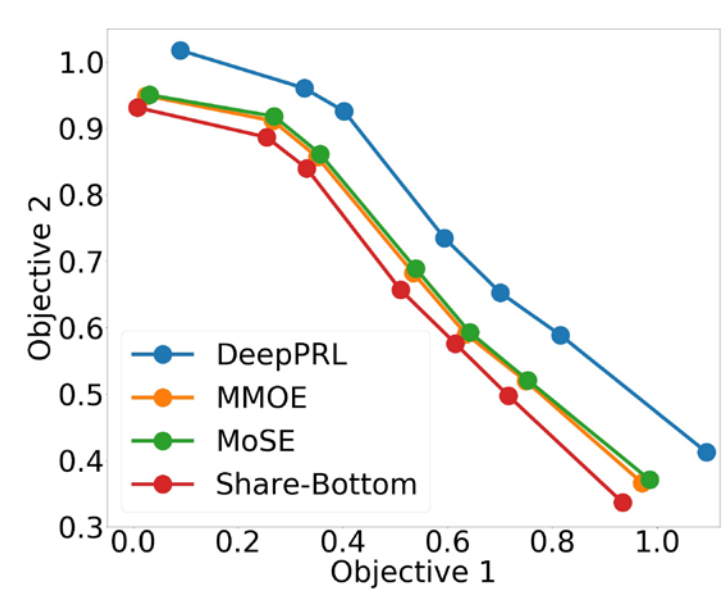

*(b) Iteration 10: +3.25%*

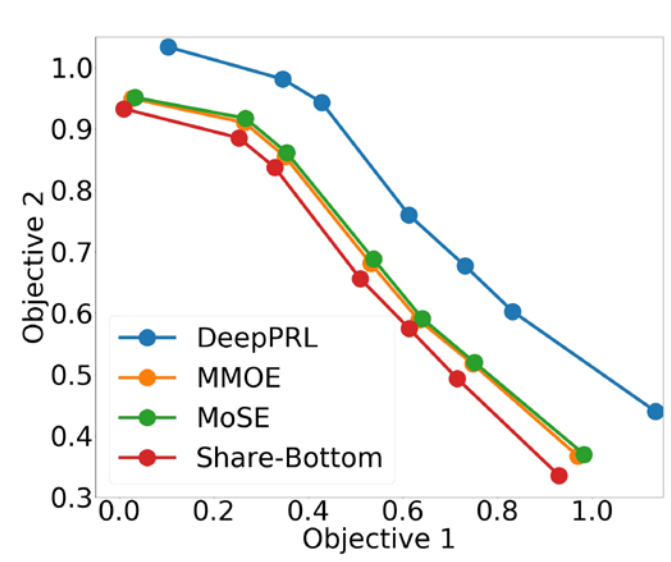

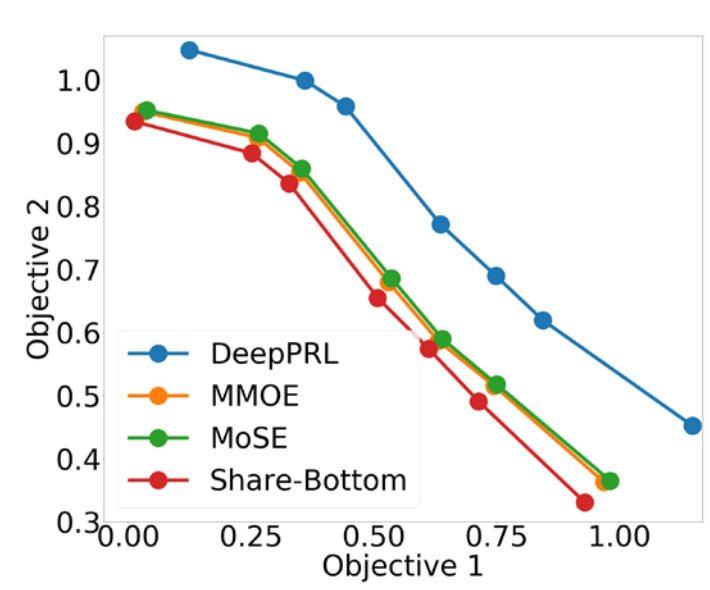

*(c) Iteration 25: +4.14%* *(d) Iteration 50: +5.20%*

*Figure 11: Long-Term Performance Analysis from the Simulation Study, Along with Improvements of Convex Hull Volume Over the Second-Best Baseline (MoSE)*

# 7 Online Experiment

## 7.1 Empirical Context and Identification

To demonstrate the business impact of our proposed model, we conduct a large-scale online controlled experiment on live customers at Alibaba-Youku, where we focus on following three objectives based on the company guidelines: *Click-Through Rate (CTR)*, a binary value measuring whether the user has clicked on and watched at least 10 seconds of the video; *Video View (VV)*, a binary value measuring whether the user has finished watching the entire video or at least 90% of the content; and Dwell Time (DT), a continuous value measuring the user watching time in seconds. The relationships between them are complex and heterogeneous: e.g., while given a fixed session length, an increase in VV typically associates with a decrease in DT (since users can finish watching shorter videos quickly), it is still possible for the DT value to increase, especially when we recommend a series of longer videos that better match with users' tastes, rather than recommending shorter videos that will only be skipped by users (Li et al. 2020). Therefore, it is an important task to achieve simultaneous improvements across all three metrics.

We randomly split consumers into the control group and the treatment group, where they receive video recommendations served by our DeepPRL model and the latest production system (a modified version of MMOE (Ma et al. 2018) with static objective weights). The randomized splitting is achieved through binary hashing on User IDs, which ensures the balance of consumer demographics/characteristics in both groups. This has been verified by the engineering team of Alibaba as well as the p-values from the Wilcoxon rank-sum test shown in Table 11, where the differences for each consumer feature between the two groups are all statistically insignificant. Therefore, we validate the randomized setting in our experiment, which enables us to estimate the treatment effect directly. In addition, consumers are unaware of their treatment assignment during the experiment as the GUI remains the same, therefore eliminating the novelty effect.

Our experiment was conducted during the whole month of January 2021 with 63,361 consumers and 30,714,995 watching records. The consumer pool and experiment duration are determined following the standard protocols at Alibaba and deemed to be sufficient to demonstrate the benefits of our model, according to the director of the experiment team. We use OLS to specify the three objectives $VV_{ijt}$, $DT_{ijt}$, and $CTR_{ijt}$ for consumer $i$ and video $j$ at time $t$:

$$Objective_{ijt} = \alpha_0 + \alpha_1 Treatment_i + \overrightarrow{\alpha_2}\overrightarrow{X_i} + \overrightarrow{\alpha_3}\overrightarrow{Y_j} + D_t + \varepsilon_{ijt} \quad (8)$$

where $Treatment_i$ is the dummy variable indicating the consumer group assignment, $\overrightarrow{X_i}$ represents consumer features, $\overrightarrow{Y_j}$ represents video features, and $D_t$ represents the time-fixed effects. We list the detailed consumer and video features used in our analysis in Table 11 below.

| Category | Variable | Description | Format | p-value |
|---|---|---|---|---|
| Consumer Features | Gender | Gender of the consumer | Categorical | >0.05 |
| | Age | Age of the consumer | Numerical | >0.05 |
| | Province | The province where the consumer lives in | Categorical | >0.05 |
| | City | The city where the consumer lives in | Categorical | >0.05 |
| | OS | The operating system on the consumer's device | Categorical | >0.05 |
| | VIP Status | Consumer subscription status to VIP | Categorical | >0.05 |
| | Active Days | The number of days that the consumer has logged into the platform over the past month | Numerical | >0.05 |
| Video Features | Genre | Genre of the video | Categorical | |
| | View Count | The total view number over the past month | Numerical | |
| | Comment Count | The total comment number over the past month | Numerical | |
| | Release Days | The number of days that the video has been released on the platform | Numerical | |
| | Video Length | The duration of the video | Numerical | |

*Table 11: Summary of Consumer and Video Features Recorded in the Online Experiment. p-values from*

*the Wilcoxon rank-sum test between the two consumer groups are shown in the last column.*

### 7.2 Main Results

We present the regression analysis in this section that directly compares consumers' video-watching behaviors between the treatment and control groups. As Table 12 demonstrates, our model significantly increases video consumption compared with the latest production model, as we observe significantly positive treatment coefficients for all three objectives. In particular, in each session, consumers in the treatment group increased the amounts of viewed videos (VV) by 0.474 (or 5.09% of the average VV value), the average watching time (DT) by 45.836 seconds (or 7.11% of the average DT value), and the average click-through rates (CTR) by 2.36%, having $p < 0.01$ in all cases. This significant improvement is due to our model's capability of balancing three conflicting objectives in a personalized and contextualized manner. In addition, we have confirmed with the company that there are no significant differences in churn rates between both groups throughout the experiment, demonstrating that improvements indeed come from the benefits of our model, rather than the abandonment effect. It also does not incur additional computational costs. As a result, it has demonstrated a very substantial economic impact, leading to a potential 20 million USD annual revenue increase based on the fiscal report 2021-2022 of the company, the analysis of the experiment team, and the fact that three metrics are directly associated with the advertisement sales and effectiveness on the video streaming platform. As a result, the company has deployed our model into production for all consumers on the platform.

| | VV | DT | CTR |
|---|---|---|---|
| Treatment | 0.4742***<br>(0.0611) | 45.836***<br>(6.6932) | 0.0236***<br>(0.0049) |
| Consumer Features | Yes | Yes | Yes |
| Video Features | Yes | Yes | Yes |
| Time Fixed Effect | Yes | Yes | Yes |
| R-Squared | 0.0078 | 0.0089 | 0.0093 |
| Consumers | 63,361 | 63,361 | 63,361 |
| Observations | 30,714,995 | 30,714,995 | 30,714,995 |

*Table 12: Average Treatment Effect of Adopting Our Proposed Model through the OLS model. The table shows robust standard errors in parentheses. *p<0.1; **p<0.05; ***p<0.01*

### 7.3 Impact on Objective Relationships and Potential Mechanisms of Performance Gain

To better understand the significant performance improvements, we conduct additional analysis in this section to study how the relationship between different objectives changes before and after the adoption of our proposed model in the online experiment. To that end, we measure the Pearson Correlation Coefficients between the three objectives in pre- and post-treatment periods among all the observed records of the treatment group. The results in Table 13 show that our model alleviates the negative correlations between VV and DT, as the relationships between them are now turning into the region of neutral or even positive. We have also conducted the same analysis all four offline datasets, where we observe a consistent positive shift of objective relationships after adopting our proposed model, since it is capable of identifying useful and personalized content for recommendation that satisfy the consumer needs across multiple dimensions simultaneously.

| Objective Pairs | Pre-Treatment | Post-Treatment |
|---|---|---|
| Online Experiment: Video View – Dwell Time | -0.163 | 0.196 |
| Online Experiment: Video View – Click-Through Rate | 0.562 | 0.651 |
| Online Experiment: Dwell Time – Click-Through Rate | 0.364 | 0.435 |
| Alibaba-Youku: Video View – Dwell Time | -0.238 | 0.003 |
| Yelp: Relevance – Novelty | -0.596 | -0.207 |
| Spotify: Listening% – Saved | 0.369 | 0.468 |
| Kuaishou: Consistency – Recency | -0.024 | 0.203 |

*Table 13: Comparison of the Objective Relationships Before and After the Adoption of Our Model*

Building on these analyzes, we further attribute the performance gain (and the positive shift of objective relationships) in our online experiment to two potential mechanisms: “longer session time” and “better recommendation match”. The former refers to the fact that users have on average spent a longer time for each session (+2% overall) after being served by our proposed model, leading to significant increases in all three metrics. This provides strong evidence that our model significantly improves user satisfaction with the video streaming platform, prompting them to spend more time in each session. The latter comes from the observation that our model still significantly outperforms the production system even after matching the session length (although the improvements are smaller due to the first mechanism). This is the case, since our model significantly improves the quality of video recommendations, so that each user will be more likely to click on the recommended videos, and that when the user does click on the video, she/he will be more likely to spend a significant amount of time on (or finish watching) the recommended video, as opposed to "click-bait" videos, resulting in a mutually beneficial objective relationships.

Finally, we conduct additional analysis of contextual factors to reveal which product segments tend to exhibit this positive shift, as we compare the objective relationships across videos with different lengths and popularities. Results presented in Figures 12(a) and 12(b) show that those videos that are less popular or are longer in terms of duration would enjoy an even greater shift of objective relationships when adopting our proposed model. These observations also match the aforementioned potential mechanisms of performance gains: since each user spends a longer session time, she/he would be more likely to click on those longer videos; and since we offer a better match in the video recommendations, we will be able to promote those less popular videos to the consumers that truly enjoy them and are willing to click and spend time on them as a result.

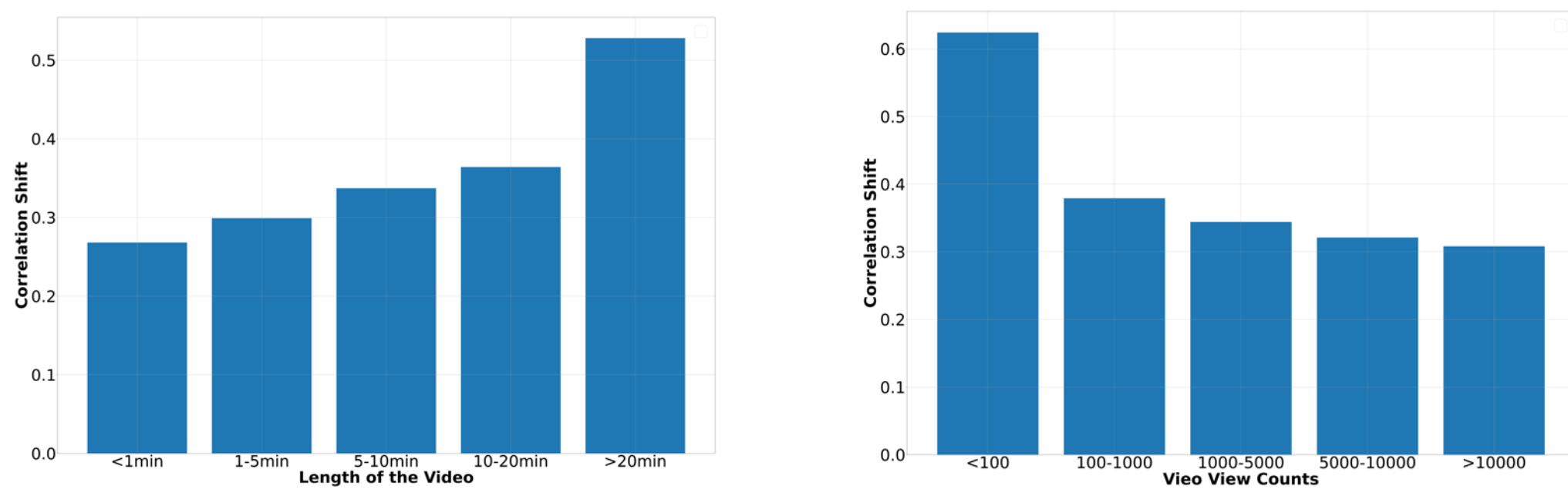

*(a) Analysis of the Length of the Video* *(b) Analysis of the View Count of the Video*

*Figure 12: Bar Plots of the Positive Shift of the Pearson Correlation Between Video View and Dwell Time, Before and After the Adoption of our Proposed Model in the Treatment Group*

### 7.4 Long-Term Analysis

As discussed in Section 4.3, an important advantage of our model lies in its ability to optimize long-term performance. To validate this point, we conduct the long-term analysis in this section by estimating the treatment effect on a daily basis following the same identification model in equation (8). We have confirmed that there is no significant difference between the two groups at any time in our experiment. Note that our one-month experiment is sufficient to demonstrate the long-term impact on the platform (in comparison, most A/B tests are conducted over one week at Alibaba, thus making our experiment relatively long-term for the company).

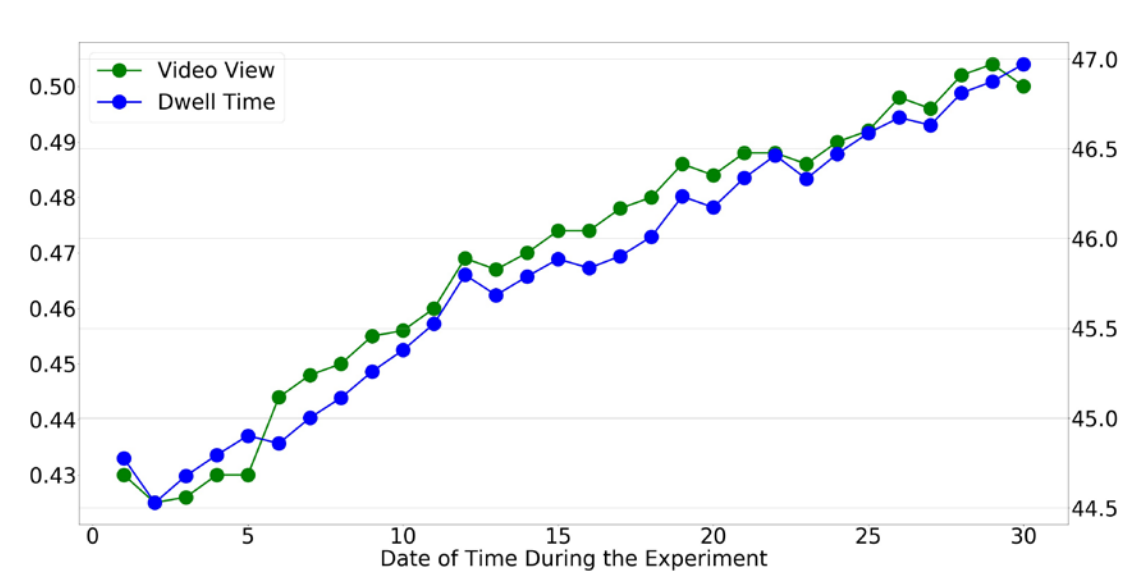

*Figure 13: Long-Term Analysis of Our Model in the Online Experiment. The Y-axis Shows the Average Treatment Effects (Performance Improvements) on Both Objectives.*

The results are presented in Figure 13, where the x-axis represents the experiment date and the y-axis denotes the treatment effects on VV and DT (we have similar observations for CTR). We observe from Figure 13 that our model significantly outperforms the production system throughout the entire experiment without any performance degradation. Furthermore, the improvements gradually increase since day 5 of our experiment, and they eventually reach the greatest level for both VV and DT by the end of our experiment. As these improvements consistently remain significant, we demonstrate the superiority of our model and its novel module for improving long-term multi-objective recommendation performance.

### 7.5 Robustness Check & Cold-Start Analysis

To further validate our findings, we also conduct an extensive set of robustness checks with the following alternative settings: (a) We include different combinations of consumer features, video features, and time-fixed effects. (b) We exclude premium video content in our analysis that is only available to VIP consumers, who might be more willing to watch those videos due to their loyalty to the platform and the exclusivity of the content. (c) We exclude those records from new consumers or newly released videos to account for the novelty or recency effects. (d) To study the capability of our model to handle the cold-start problem, we focus on those new consumers described in (c) instead and identify how the performance improvements on them will fluctuate across a certain period of time. As shown in Tables 14 and 15, our results are not affected by the exclusion of feature information, fixed effects, premium video content, or new consumer/video records. Additionally, as we show in Figure 14, our model consistently outperforms the latest production system for new consumers, and the improvements become increasingly larger as time goes by, indicating its effectiveness in handling the cold-start recommendation scenario. To summarize, our findings remain consistent across multiple alternative settings, which further demonstrates the advantages of our model and its tangible economic impact for the industry.

| | VV | VV | VV | VV | VV |
|---|---|---|---|---|---|
| Treatment | 0.4742*** (0.0611) | 0.4722*** (0.0697) | 0.4756*** (0.0648) | 0.4744*** (0.0627) | 0.4731*** (0.0749) |
| Consumer Features | Yes | No | Yes | Yes | No |
| Video Features | Yes | Yes | No | Yes | No |
| Time Fixed Effect | Yes | Yes | Yes | No | No |
| R-Squared | 0.0078 | 0.0041 | 0.0054 | 0.0071 | 0.0019 |
| Consumers | 63,361 | 63,361 | 63,361 | 63,361 | 63,361 |
| Observations | 30,714,995 | 30,714,995 | 30,714,995 | 30,714,995 | 30,714,995 |

*Table 14: Average Treatment Effect on VV with Different Combinations of Consumer Features, Video Features, and Time Fixed Effect. Similar Observations in DT and CTR. Robust standard errors in parentheses. *p<0.1; **p<0.05; ***p<0.01*

| | VV (b) | DT (b) | CTR (b) | VV (c) | DT (c) | CTR (c) |
|---|---|---|---|---|---|---|
| Treatment | 0.4742*** (0.0611) | 45.779*** (6.7015) | 0.0221*** (0.0070) | 0.4706*** (0.0650) | 45.533*** (6.9263) | 0.0224*** (0.0071) |
| Consumer Features | Yes | Yes | Yes | Yes | Yes | Yes |
| Video Features | Yes | Yes | Yes | Yes | Yes | Yes |
| Time Fixed Effect | Yes | Yes | Yes | Yes | Yes | Yes |
| R-Squared | 0.0078 | 0.0089 | 0.0093 | 0.0077 | 0.0087 | 0.0092 |
| Consumers | 63,361 | 63,361 | 63,361 | 62,378 | 62,378 | 62,378 |
| Observations | 28,378,372 | 28,378,372 | 28,378,372 | 30,538,268 | 30,538,268 | 30,538,268 |

*Table 15: Average Treatment Effect when Excluding Premium Video Content, New Consumers, and Newly Released Videos. Robust standard errors in parentheses. *p<0.1; **p<0.05; ***p<0.01*

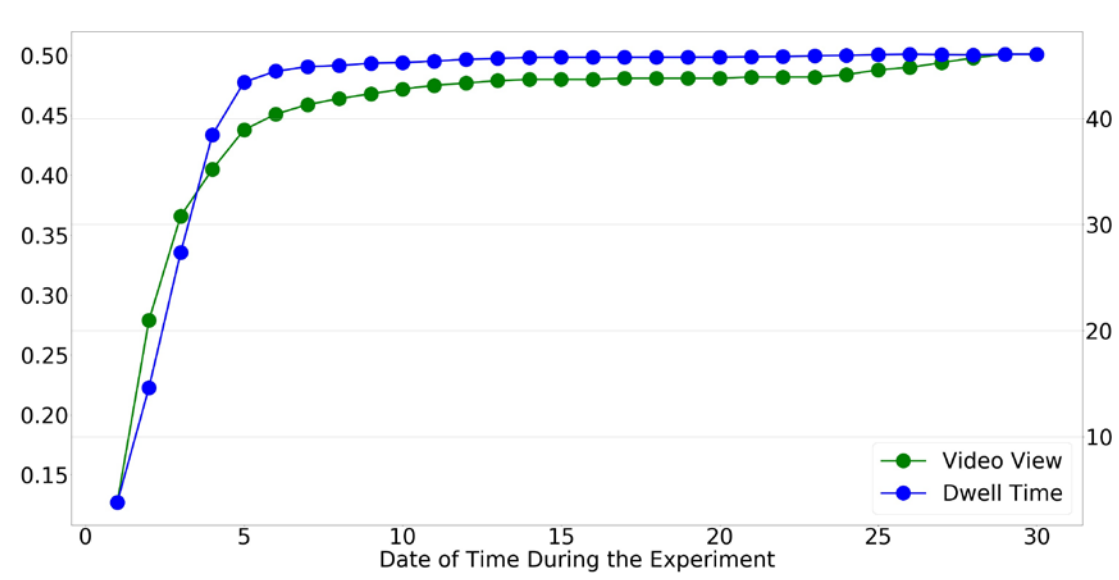

*Figure 14: Long-Term Recommendation Performance on Cold-Start Consumers*

# 8 Conclusions

## 8.1 Summary of Technical Novelties

In this paper, we propose a novel DeepPRL method to properly balance between different objectives and to improve their performance along the Pareto frontier. By taking into account the relationships between different objectives and determining the weight of each objective *in a personalized and contextualized manner*, our method provides important technical novelties and modeling benefits: First, it effectively models complex relationships between objectives by modifying HyperNetwork (Ha et al. 2016) into a multi-objective predicting network, where we incorporate a mixture-attention mechanism to capture the heterogeneous objective relationships, leading to significantly better prediction performance. Second, it enhances the level of personalization and contextualization by selecting the appropriate set of objective weights for *each* consumer under *each* contextual scenario with our Contextual Policy Gradient method. Third, by utilizing deep reinforcement learning techniques to build an end-to-end multiple objective recommendation framework, it mitigates the data sparsity and scalability problems, since it is capable of automatically and effectively identifying consumer preferences towards different objectives from archival records without requiring any explicit consumer feedback. It also enables us to optimize the long-term (versus only short-term) recommendation performance, which is crucial for the business success of the company in the long run. Finally, it is particularly flexible in the sense that it works for an arbitrary number of objectives, and that it manages to achieve significant improvements across *all* the objectives regardless of their relationships.

These benefits, as well as theoretical properties that we discussed in Section 4.4, are empirically illustrated through offline experiments on four applications, where DeepPRL outperforms state-of-the-art baselines and dominates their Pareto frontiers. Our method also maximizes the long-term performance, thus obtaining even greater performance improvements vis-à-vis existing solutions. Finally, in the online experiment, our method achieved significant improvements over the latest production system across multiple conflicting objectives in the long run, illustrating its tangible economic impact on the industry.

**8.2 Managerial Implications**

For platforms, optimizing solely for one metric, like relevance, can lead to undesirable consequences discussed in Section 1. Therefore, managers shift from seeking a dominating objective to identifying several objectives and strategically selecting the most preferable compromise at the Pareto Frontier, i.e., the solution that best aligns with the organization's strategic priorities and risk preferences. With DeepPRL, we achieve superior performance sufficiently close to the Pareto optimality, since we effectively balance different objectives regardless of their relationships. This should foster deeper user trust and loyalty as compared to systems narrowly focused on a single objective without balancing different user needs.

Moreover, in Table 16, we illustrate four pairs of conflicting objectives across different use cases that need to be properly balanced, so that stakeholders can make better strategic decisions with respect to competing goals or inherent conflicts. For example, media platforms need to balance between accuracy and novelty to provide suitable content for each consumer, while social media platforms are obliged to incorporate ethical considerations and user engagement metrics when designing the content recommendation algorithms. With the help of DeepPRL, these objective tradeoffs will be handled more effectively to create more business value.

Finally, as we provide empirical evidence to demonstrate the importance of context and dynamic factors in the recommender system design, it implies that managers must champion a customer-centric approach, investing in understanding their needs and how context influences them, in order to adjust their business strategies and better connect with the customers.

| Objective Category 1 | Objective Category 2 | Managerial Implication |
|---|---|---|
| Accuracy, Relevance | Diversity, Novelty, Unexpectedness | Over-emphasizing the former risks filter bubbles & user boredom; under-emphasizing risks irrelevance & frustration. |
| Short-Term User Engagement | Long-term User Retention | Aggressively optimizing for short-term metrics might alienate users in the long run if they feel manipulated or if the recommendation quality declines. |
| User Satisfaction, Trust | Provider Fairness/Exposure | Focusing solely on satisfaction might lead to a "rich-get-richer" scenario where popular content dominates, while less popular but potentially relevant content is never discovered. |
| Ethical Considerations (e.g., mitigating bias & echo chambers) | User Engagement Metrics | Content that is polarizing, controversial, or even harmful can drive very high engagement. Prioritizing ethical considerations might necessitate deliberately filtering or not recommending such content, potentially at the cost of raw engagement metrics. |

*Table 16: Illustrations of Competing Objectives and Their Managerial Implications*

### 8.3 Future Work

We plan to further study the benefits of our method across different platforms and context, especially when there exists a strategic trade-off between different objectives. We are also interested in exploring methods for producing multi-objective recommendations under certain constraints. Finally, since we are dealing with the dynamic state transitions of objective weights, it is worth studying the potential of a state-space model to solve the optimization problem.